\documentclass[leqno]{amsart}

\usepackage[foot]{amsaddr}

\usepackage{color}
\usepackage{geometry}

\usepackage[numbers]{natbib}
\usepackage{amsmath}
\usepackage{mathtools}

\usepackage{subfiles}
\usepackage{graphicx}
\usepackage[section]{placeins}
\usepackage{hyperref}
\usepackage{stmaryrd}
\usepackage{amsmath,amssymb,amsfonts,amsthm,textcomp}
\usepackage{mathtools}
\usepackage{tensor}
\usepackage{epstopdf}
\usepackage{multicol}
\usepackage{empheq}
\usepackage{mathrsfs} 
\usepackage{tikz,pgfplots}
\usetikzlibrary{calc}
\usepackage{enumitem}

\usepackage[linesnumbered,ruled,vlined]{algorithm2e}
\usepackage{subfigure}
\usepackage{lscape}
 \usepackage{tabularx} 
 \usepackage{booktabs}
\usepackage{fancyhdr}
\graphicspath{{./Figures/}
              {./Figures/Base displacement/} 
              {./Figures/Final Displacement/}
              {./Figures/Max displacement/}
              {./Figures/Displ history/}
			 }

\newfont{\tenbfsl}{cmbxti9 scaled 1200}
\newfont{\tenbbb}{msbm10}
\newfont{\svnbbb}{msbm8}

\theoremstyle{remark}

\theoremstyle{definition}

\newcounter{syn}[section] 
\renewcommand{\thesyn}{\arabic{section}.\arabic{syn}}

\makeatletter
\def\threevdots{\mskip+4mu\vbox{\baselineskip2.25\p@ \lineskiplimit\z@
  \kern4.9\p@\hbox{.}\hbox{.}\hbox{.}}\mskip+3.8mu}
\makeatother

\begin{document}

\title[A high-fidelity seismic intensity measure to assess dynamic liquefaction in tailings]{A high-fidelity seismic intensity measure to assess dynamic liquefaction in tailings}
\author{Nicol\'as A. Labanda $\spadesuit$}
\address{$\spadesuit$ SRK Consulting (Australasia) Pty Ltd, Perth, WA, Australia}
\email{nlabanda@facet.unt.edu.ar (N.A. Labanda)}
\author{Roberto J. Cier $\spadesuit$}
\email{rcier93@gmail.com (R. Cier)}
\author{Mauro G. Sottile $\flat$}
\address{$\flat$ SRK Consulting, Argentina}
\email{msottile@srk.com.ar (M. Sottile)}

\date{\today}

 \begin{abstract}
 \noindent

Deformation analyses of tailings dams under dynamic conditions require using earthquake records as input loading. Moreover, these records must represent the local seismicity, expressed by ground motion power indicators denominated intensity measures (IM). The ability and accuracy to describe the characteristics of a seismic record play a fundamental role in earthquake engineering and damage assessment of geotechnical facilities. None of the existing IMs represents a robust enough predictor of a given seismic demand (e.g., residual displacements). Different signals may generate a wide spectrum of results, with diverse effects that could produce insignificant damage to global failure depending on the structure. Usual engineering procedures select a huge number of records to overcome this limitation and develop a large set of numerical simulations to bound the uncertainty of the results, which becomes a time-consuming approach. This paper presents a new high-fidelity seismic IM to perform more accurate ground motion {selection}, which captures the spectral properties of the record for the frequency content that the dam does not filter. This IM represents a way to estimate beforehand a seismic demand, expressed, for instance, in terms of displacements. The proposed IM is applied to a finite element model for an upstream tailings dam cross-section, using a constitutive model capable of capturing dynamic liquefaction. The obtained results show that our proposal gives highly reliable correlations with different selected demands. Comparisons with classical IMs are also discussed, showing that our proposal emerges as a practical solution to a large dated discussion within our community.


 \end{abstract}

\maketitle


\tableofcontents                        


\section{Introduction}
\label{section:Introduction}

Selecting a set of ground motions representative of the local seismic condition is decisive when assessing dynamic liquefaction. Seismologists usually collect several thousand seismic records, and after performing deterministic and probabilistic seismic hazard analyses, they pick a small portion of them to use in analytical or numerical models. Most of the available strategies, definitions and procedures used to that end are borrowed from structural engineering, being those suitable when the analysed structure is, for instance, a concrete building with a well-known elastic regime and, therefore, with a natural period that can be computed. However, if the analysis is developed for soft-ground-made geotechnical structures such as tailings storage facilities (TSF), where the elastic regime is almost negligible, these techniques provide inaccurate outcomes. In that sense, the authors believe that new definitions suitable for geotechnical purposes represent an improvement opportunity and must be developed by our community to overcome those well-known limitations with rather poor correlations between IMs and actual structural demand.

Over the last decades, researchers proposed several IMs characterise the destructive potential of a seismic record: peak ground velocity (PGV) and peak ground acceleration (PGA), the two most widespread but highly limited of them~\citep{Ebrahimian2012}; Arias intensity (AI), which offers a notion of the total energy content of the signal~\citep{Arias1990}; modified cumulative absolute velocity measures such as CAV and CAV5, being the latter the integral of the acceleration after applying a 5 $\text{cm}/\text{s}^2$ acceleration threshold \citep{Kramer2006}; normalised hysteretic energy, an empirical relation between dissipated shear energy and residual excess pore pressure ratio~\citep{Green2000}; among others. 

Taking advantage of its simplistic and computationally efficient formulation, some researchers have used the sliding-block Newmark-type models~\citep{Newmark65} to estimate settlements of slopes under shaking and adopted empirical modifications of the Arias intensity to characterise the excitation~\citep{CHOUSIANITIS2014}. Combinations of Newmark-type models as a displacement estimator and some selected IMs such as PGA, CAV and AI are popular in hazard analysis among geotechnics~\citep{Jibson1993521, JIBSON2007209,Armstrong2013,Roy2016,DEYANOVA2016210}. Nonetheless, due to its constitutive limitations, the Newmark sliding block model has a limited prediction capability to assess dynamic liquefaction in soft soils and tailings.

Contributions related to dynamics/liquefiable soils are limited, although new approaches have been proposed during the last decade~\cite{Sottile2021}. Naeini et al.~\cite{NAEINI2018179} addressed the problem of tailings dams subjected to dynamic loads by focusing on detecting resonance points employing transfer functions~\citep{Severn99,Hwang2007}. Kramer et al.~\cite{Kramer2016} reviewed procedures to detect the time of liquefaction triggering, comparing their performance with empirical methods. The research is focused on signal analysis using {short-term} Fourier transform (STFT), spectrograms, wavelets transforms and Stockwell spectrum procedures, showing that the mean frequency content tends to reduce in signals recorded above a liquefied stratum \citep{Kramer2018384,MEZAFAJARDO2019292,ozener2020}. With the continuous computational power enhancement, numerical models have become the standard procedure to perform industry and forensic studies over real dam failures, with remarkable works \citep{ISHIHARA20153,BAOTIAN14,Wenlian12,swidqs2016}. Despite advances in this subject matter, tailings engineering keeps using correlations of engineering demand parameters with classical IMs, usually leading to highly scattered results \citep{HARIRIARDEBILI2019761}. Consequently, Ref. \cite{LABANDA2021106750} showed that computing the spectral properties of the seismic signals in low frequencies can accurately predict the liquefaction triggering in tailings dams under strong ground motions when demand is simulated using numerical and analytical methods.

Despite all the remarkable contributions proposed by the geotechnical community, there is still a lack of accuracy in the predictive capacity of available tools, mainly when the ground motion produces highly non-linear soil behaviour. This paper presents a high-fidelity seismic intensity measure based on spectral features of the seismic input record suitable to assess tailings dam failure. This new IM is used to produce a preliminary estimate of the seismic demand of any given set of earthquakes. The proposal is validated by selecting the most demanding earthquakes and showing that this method correlates well with the displacement contours obtained in dynamic numerical models for an upstream TSF undergoing dynamic liquefaction. This paper is written to complement other contributions published by the authors, now focused on assessing the accuracy of spectral approaches for mid-to-low intensity seismic records.

The paper is organised as follows: Section \ Ref {section:insight} defines the proposed intensity measure based on seismic spectral properties, which is then applied to a set of seismic records used for the dynamic liquefaction analyses. Section \ Ref {section: non-linear} describes the finite element model used to assess the proposal, including constitutive model parameters, boundary conditions and staged construction to raise the tailings dam. Section \ Ref {section:Examples} shows the proposed tool's predictive capability and efficiency, showing that correlations with the demand are far better than those obtained with classical approaches like Arias intensity and PGA. Finally, some conclusions and outlooks are drawn in Section \ref{section:Conclusions}.

\section{Seismic intensity measure}
\label{section:insight}

This section summarizes a new intensity measure based on spectral features and then is applied to a set of mid-to-low intensity seismic signals and compared with other classical quantities.

\subsection{Definition}

The proposed intensity measure is computed using spectral decomposition using the Fast Fourier Transform (FFT), an efficient implementation of the Discrete Fourier transform (DFT). In this sense, each seismic record is decomposed into trigonometrical functions with different frequency contents and amplitudes, which allows computing the spectral power for each discrete frequency. The formulation is stated in a discrete space, assuming a discrete seismic signal $\left\lbrace \boldsymbol{a}_n \right\rbrace = a_0, a_1, ..., a_{N-1}$ expressed as a finite set of $N$ elements uniformly spaced of time-history accelerations. The DFT of the seismic records is defined employing Euler's formula:
\begin{equation}
	\begin{split}
		\mathcal{F}\left\lbrace \boldsymbol{a}_n \right\rbrace \left( j \right) = \left\lbrace \boldsymbol{A}_j \right\rbrace &= \sum^{N-1}_{n=0} a_n \cdot e^{-i2 \pi j\frac{n}{N}} = \sum^{N-1}_{n=0} a_n \cdot \left[\cos\left(\frac{2 \pi}{N} j n\right) + i \cdot \sin\left(\frac{2 \pi}{N} j n\right) \right]  
	\end{split}
	\label{eq:DFT}
\end{equation} 
where $ \left\lbrace  \boldsymbol{A}_j \right\rbrace$ is a set of complex vectors representing the amplitude and phase of a complex sinusoidal component and $j$  an integer representing the frequency domain. The power spectrum density in terms of the frequency is defined as the norm of the amplitude in each discrete frequency
\begin{equation}
	\mathcal{S} \left( j \right) = \| \mathcal{F} \left\lbrace \boldsymbol{a}_n \right\rbrace \left( j \right) \|^2  
	\label{eq:PSD}
\end{equation} 

Then, the intensity measure defined in terms of the spectral power of the signal in all the frequency windows is
\begin{equation}
	P_{0 - \infty} = \sum^{\infty}_{j=0} \mathcal{S}\left( j \right) \Delta j  
	\label{eq:PS}
\end{equation} 
where $\Delta j$ is the frequency sampling. As was proved in other contributions \cite{Kramer2016,LABANDA2021106750}, soil liquefaction is sensitive to the energy content in low frequencies. For practical purposes, the seismic intensity measure used is a windowed version of the spectral power expressed in equation \eqref{eq:PS}, where the highest frequency considered for the calculations is the limit of power accumulation, i.e., $P_{0 - X \ Hz}$ is the accumulated power between the frequencies 0 to  $X \ Hz$. 
Applying Parseval's Theorem to the discrete signal, we obtain
\begin{equation}
	\sum^{N-1}_{n=0} \|  a_n \|^2 = \sum^{\infty}_{j=0} \|   \mathcal{F} \left( j \right) \|^2  
	\label{eq:parseval}
\end{equation} 
which means that when the amplitude is integrated on its full frequency domain, the proposed intensity measure expressed in terms of the spectral decomposition tends to represent classical intensity measures based on the integration of the seismic signal, such as the Arias intensity. In this way, this spectral power approach presented here is stated as a generalization of these classical IMs. The spectral power is usually expressed in an adimensional manner to be plotted using a so-called spectrogram, which shows the signal decomposition in terms of time, frequency and spectral power. Then, the spectral power expressed in decibels $dB$ can be computed as
\begin{equation}
	P_{dB} = 10 \log_{10} \left( \frac{P}{P_{r}}\right)  
	\label{eq:dbpower}
\end{equation}
where $P$ is the computed spectral power and $P_r = 10^{1.5}$ is a reference power. The reference power behaves like a simple shift in the accumulated power and does not modify the proposed intensity measure.

\subsection{Characterization of seismic records}

A set of 14 records of site-representative seismic signals were scaled using records from the PEER NGA-West2 database and following hazard curves obtained from a seismic hazard analysis using a recurrence period of 50,000 years. Then, the PGA of the original records was scaled to $0.213$ g or $2.094 \frac{\text{m}}{\text{s}^2}$, and all of them correspond to seismographs located on dense soil/soft rock (NEHRP site class C). This topic is out of the scope of this paper, and we will limit our discussion to show the novelty of our approach to predicting the most demanding record.

Many signals available include long periods of near-zero acceleration recordings, unnecessarily impacting the computation time when running numerical models. As part of these works, records are processed following the PEER procedure \cite{Ancheta2013}. This methodology consists of low and high-pass Butterworth filters applied in the frequency domain, applying a simple baseline correction for those cases where filtering did not remove non-physical trends in the displacement time series. In addition, signals are truncated such that the remaining energy is 99$\%$ of the original signal.

The main characteristics of scaled records are summarized in Table \ref{T:SeismicRec}. The spectral power of four frequency windows is included together with arias intensity (AI). Spectral power is expressed in relative terms, i.e. the argument of the logarithm,  as explained in equation \eqref{eq:dbpower}. Those 14 selected records correspond to seven events, where signals in east-west (EW) and north-south (NS) directions were adopted. If we follow the Arias intensity as characteristic of the structure's demand, we would wrongly conclude that the Tottori event in EW and NS directions are the stronger ground motions to be used. 

As a counterpart, when our intensity measure is used, we can predict that the most demanding record will be Chi-Chi Taiwan 05 (TCU064) NS and EW, along with Chi-Chi Taiwan 05 (TCU076) EW. In the following section, we will show that this last prediction is the right one, while the Tottori earthquake ranks as an average demanding event as it is also well predicted by our IM. 

%

\begin{table}
	\centering
	\caption{Seismic records used for the dynamic liquefaction analysis.}
	\label{T:SeismicRec}
	\begin{small}
		\begin{tabular}[t]{lccccccccccccc}
			\toprule
			Label & Event Name & Direction & Duration  & AI  & $\frac{P_{0-1.5}}{P_r}$ & $\frac{P_{0-2.0}}{P_r}$ & $\frac{P_{0-2.5}}{P_r}$ & $\frac{P_{0-3.0}}{P_r}$  \\
			&    &   & $\left[sec \right]$  & $\left[m/sec\right]$ & $\left[-\right]$ & $\left[-\right]$ & $\left[-\right]$ & $\left[-\right]$ \\
			\midrule
			1A & Northridge & EW	& 27.98	&	7.086 & 2608 & 3717 & 	5064 & 6511 \\
			1B & Northridge & NS	& 27.98	&	4.515 & 1595 & 2187 & 2808 &	3357  \\
			2A & Chi-Chi Taiwan 05 (TCU064) & EW	& 66.99  &	8.716 &  4930  &6685 & 8491 & 10336 \\
			2B & Chi-Chi Taiwan 05 (TCU064) & NS	& 66.99	&	9.620 &  6797  &	9277 & 	11803  & 	14295
			\\
			3A & Chi-Chi Taiwan 05 (TCU082) & EW	& 70.995	 &	5.301 &  1168	& 1797 & 2581 &	3463
			\\
			3B & Chi-Chi Taiwan 05 (TCU082) & NS & 70.995 &	8.013 &  2121	& 3099 & 4207 & 5365
			\\
			4A & Chi-Chi Taiwan 06 (TCU076) & EW & 71.995  &	7.100 &  4571	& 6193 & 7754 & 9164
			\\
			4B & Chi-Chi Taiwan 06 (TCU076) & NS & 71.995 &	6.067 & 1907 &	2618 &	3365 & 	4120
			\\
			5A & Tottori Japan & EW & 251.995	 &	11.633 &  2127 & 2895 & 3678	 & 4445
			\\
			5B & Tottori Japan & NS & 251.995 &	11.558 &  3162 & 4236 &	 5324	& 6422
			\\
			6A & Chetsu-oki & EW & 183.99 &	8.955 &  3125	& 4174 & 5284 & 6286
			\\
			6B & Chetsu-oki & NS & 183.99 &	8.045 &  2796 & 3918 &	5159 & 6211
			
			\\
			7A & Christchurch  & EW & 47.98	 &	4.001 &  915 & 1404 & 2027 & 2671
			\\
			7B & Christchurch  & NS & 47.98 &	4.630 & 581 & 848 &	1179 & 	1533
			
			\\
			\bottomrule
		\end{tabular}
	\end{small}
\end{table}
\section{Numerical model}
\label{section:nonlinear}

\subsection{Simulation layout}
The analyses aim to estimate the earthquake-induced deformations and assess the TSFs vulnerability to liquefaction under seismic loading, computing the demand at two points of the slope, one on the crest and the other on the base. For the dynamic analysis itself, the following procedure was carried out:
\begin{itemize}
	\item[•] Dynamic deformation analyses using PM4Sand model for accounting for the potential of induced tailings liquefaction.
	\item[•] Time evolution of the displacement fields and excess pore pressures during and after each seismic event.
	\item[•] Assessment of a potential failure surface due to seismically induced tailings liquefaction.
\end{itemize}

\begin{figure}[h!]
	{\includegraphics[height=5cm]{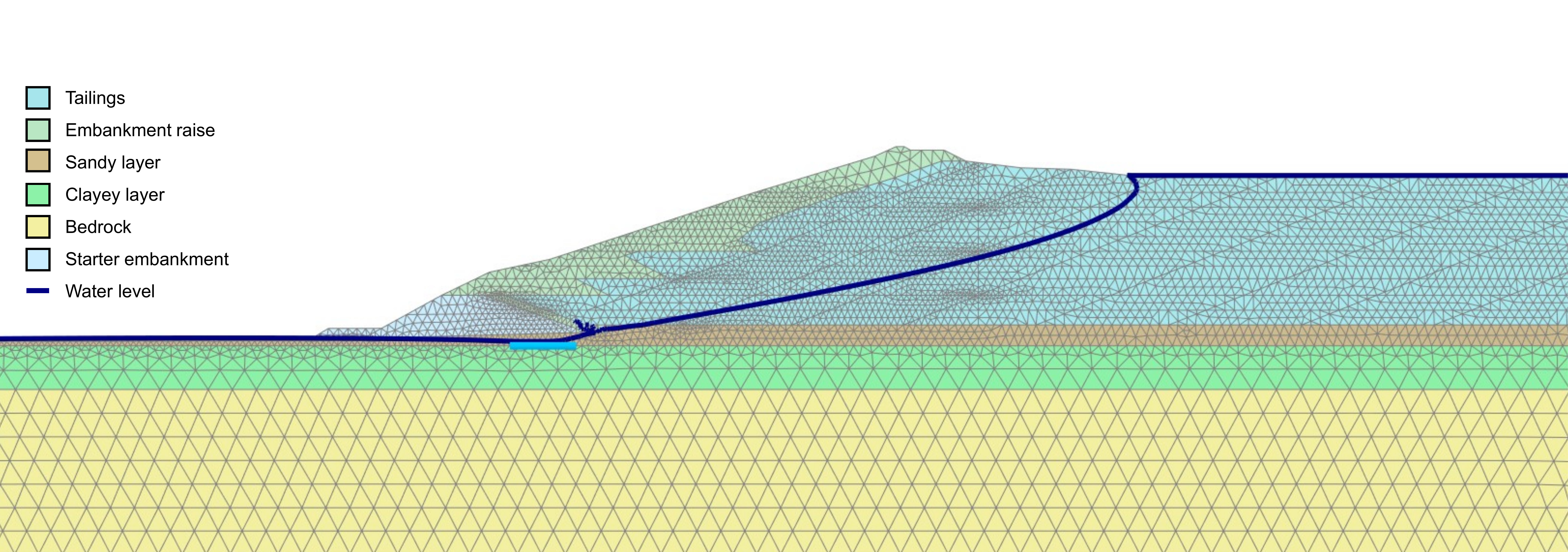}}
	\caption{Finite element mesh and materials}
	\label{fig:MeshMat}
\end{figure}

The PM4Sand constitutive model is used to simulate the dynamic behaviour of the tailings. The reader can find a detailed explanation of these constitutive models in~\cite{Boulanger2017}. Standard finite element plane-strain analyses are performed, where the TSF mesh and materials are shown in Figure~\ref{fig:MeshMat}. The full interaction between the ground, materials and pore water is simulated by employing finite element technologies capable of simultaneously reproducing the behaviour of the solid and fluid phases of the various materials that make up the TSF and its foundation. The model uses 6-node triangular elements for the staged construction and 6-node elements for the dynamic stages, as these are more stable numerically for this type of analysis. The mesh size has been confirmed to be appropriate for dynamic modelling, following recommendations in Ref.~\cite{Lysmer1969}.

\begin{figure}[h!]
	{\includegraphics[height=10cm]{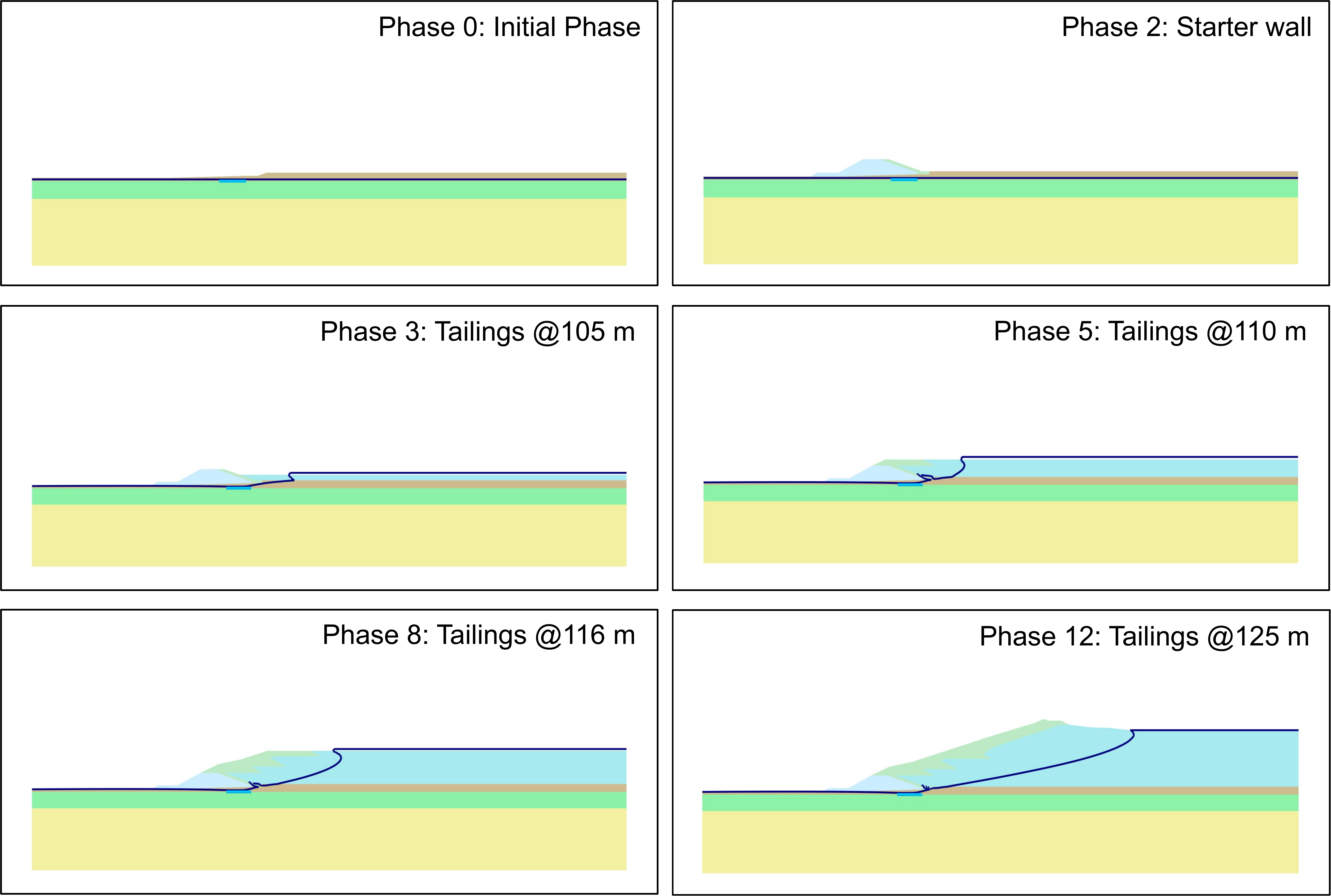}}
	\caption{Staged construction.}
	\label{fig:StageConst}
\end{figure}

Figure \ref{fig:StageConst} displays several stages of the TSF raising sequence. We computed steady-state seepage flow for each construction phase. The staged construction simulation of the facility is summarised as follows:
\begin{itemize}
	\item[•] Computation of initial stress state at the foundation;
	\item[•] Construction of the starter embankment;
	\item[•] Tailings raise using a rate in agreement with real procedures;
	\item[•] The dam's raising was done using the Hardening Soil Small model (HS-small), and tailings were switched to PM4Sand right before the dynamic phase;
\end{itemize}

Free-field boundary conditions are used on the left and right ends of the model. At the bottom, a compliant base boundary condition is employed, where the acceleration-time signals are input. According to the standard procedure for compliant bases, the input signal must be 50$\%$ of the acceleration recorded in time histories to consider that the outcrop motion is characterised by the upward incoming and downward reflected waves.

\subsection{Constitutive models for staged construction}

Tailings are mostly saturated and loose with a clear contractive behaviour under static and dynamic undrained loading, producing stain softening due to the generation of excess pore pressures. Consequently, chosen constitutive models must be able to represent this kind of quasi-brittle behaviour in the numerical simulation. The deformation analysis presented in this document uses the following constitutive models:
\begin{itemize}
	\item[•] Hardening Soil with Small Strain Stiffness (HSS): used to simulate the mechanical behaviour of tailings and non-tailings materials during the staged construction until its final elevation. This model is chosen based on its capability to reproduce various stress paths in drained and undrained conditions with satisfactory outcomes for a wide spectrum of geomaterials. 
	\item[•] Plasticity Model for Sand (PM4Sand): used to simulate tailings material behaviour during the dynamic deformation modelling. The model is chosen due to its ability to reproduce the excess pore pressure generation under cyclic loading and accurately reproduce the N-CSR curves for different relative densities of the soil. 
	\item[•] Permeabilities in coupled phases are considered different for the horizontal direction $k_h$ and the vertical $k_v$, where the latter is usually lower due to the deposition method.
\end{itemize}

Table \ref{t:HSmall} for each geotechnical unit shows the HS-small model parameters. For tailings material, some parameters are adjusted to reproduce the strain-softening with compatible peak/residual shear strength ratios, as is explained in \cite{Sottile2019}.
\begin{table}[h!]
	\caption{{HSS} material parameters used in static stages.}
	\label{t:HSmall}
	\centering
	\begin{tabular}{p{1.5cm}p{1.5cm}p{1.5cm}p{1.5cm}p{1.5cm}p{2cm}p{2.0cm}p{1.5cm}}
		\hline
		  Parameter              & Units       &     {Tailings}     &    Sandy layer     &    Clayey layer    &  Embankment raise  & Starter \mbox{embankment} &      Bedrock       \\ \hline
		$\gamma_{sat}$  & {$kN/m^3$}   &         22         &         20         &         19         &         20         &         18         &         20         \\
		$\phi'$         & $^{\circ}$ &         40         &         30         &         25         &         24         &         32         &         40         \\
		$c'$            & kPa      &         1          &         0          &         0          &         27         &         5          &        100         \\
		$\psi$          & $^{\circ}$ &         0          &         0          &         0          &         0          &         0          &         0          \\
		$G^{ref}_0$     & MPa      &         55         &        100         &         75         &         70         &         70         &         -          \\
		$\gamma_{0.7}$  & -          &     $10^{-4}$      &     $10^{-4}$      &     $10^{-4}$      &     $10^{-4}$      &     $10^{-4}$      &     $10^{-4}$      \\
		$E^{ref}_{ur}$  & MPa      &         60         &         80         &         60         &         24         &         24         &         -          \\
		$E^{ref}_{50}$  & MPa     &        6.5         &         26         &         20         &         10         &         10         &        200         \\
		$E^{ref}_{oed}$ & MPa      &         9          &         26         &         20         &         10         &         10         &         -          \\
		$m$             & -          &        0.80        &        0.50        &        0.50        &        0.75        &        0.75        &         -          \\
		$\nu_{ur}$      & -          &        0.3         &        0.2         &        0.2         &        0.2         &        0.2         &        0.3         \\
		$K^{nc}_{0}$    & -          &        0.60        &        0.50        &        0.57        &        0.59        &        0.47        &         -          \\
		$k_h$           & m/day      & $3.0\times10^{-3}$ & $8.6\times10^{-1}$ & $4.3\times10^{-2}$ & $4.3\times10^{-3}$ & $4.3\times10^{-3}$ & $4.3\times10^{-2}$ \\
		$k_v$           & m/day      & $4.3\times10^{-3}$ & $1.7\times10^{-1}$ & $4.3\times10^{-3}$ & $8.6\times10^{-4}$ & $8.6\times10^{-4}$ & $1.7\times10^{-3}$ \\ \hline
	\end{tabular}
\end{table}

\begin{table}[h!]
	\caption{PM4Sand {model} parameters for tailings, used in dynamic stages.}
	\label{t:PM4sand}
		\centering
		\begin{tabular}{llc}\hline
			Parameter & Units & Parameters\\\hline
			$\gamma_{sat}$  & $kN/m^3$ & 22.0 \\
			$G_0$     &         -        &  550    \\
			$h_{p0}$ &  -    &  15 \\
			$p_{ref}$&  kPa    &  103 \\
			$e_{max}$& -    & 1.05 \\
			$e_{min}$&  -   & 0.35 \\
			$n^b$  &  -     &  0.50 \\
			$n^d$ & -  & 0.20 \\
			$\phi'_{cv}$ & $^{\circ}$  & 33 \\
			$\nu$  & -  & 0.30 \\
			$Q$ & -  & 10.0 \\
			$R$ & -  & 2.4 \\
			$D_r$ & -  & Variable with $\psi = 0.1$ \\
			\hline
		\end{tabular}
\end{table}

\subsection{Dynamic stage set-up}

\begin{figure}[h!]
	{\includegraphics[height=7cm]{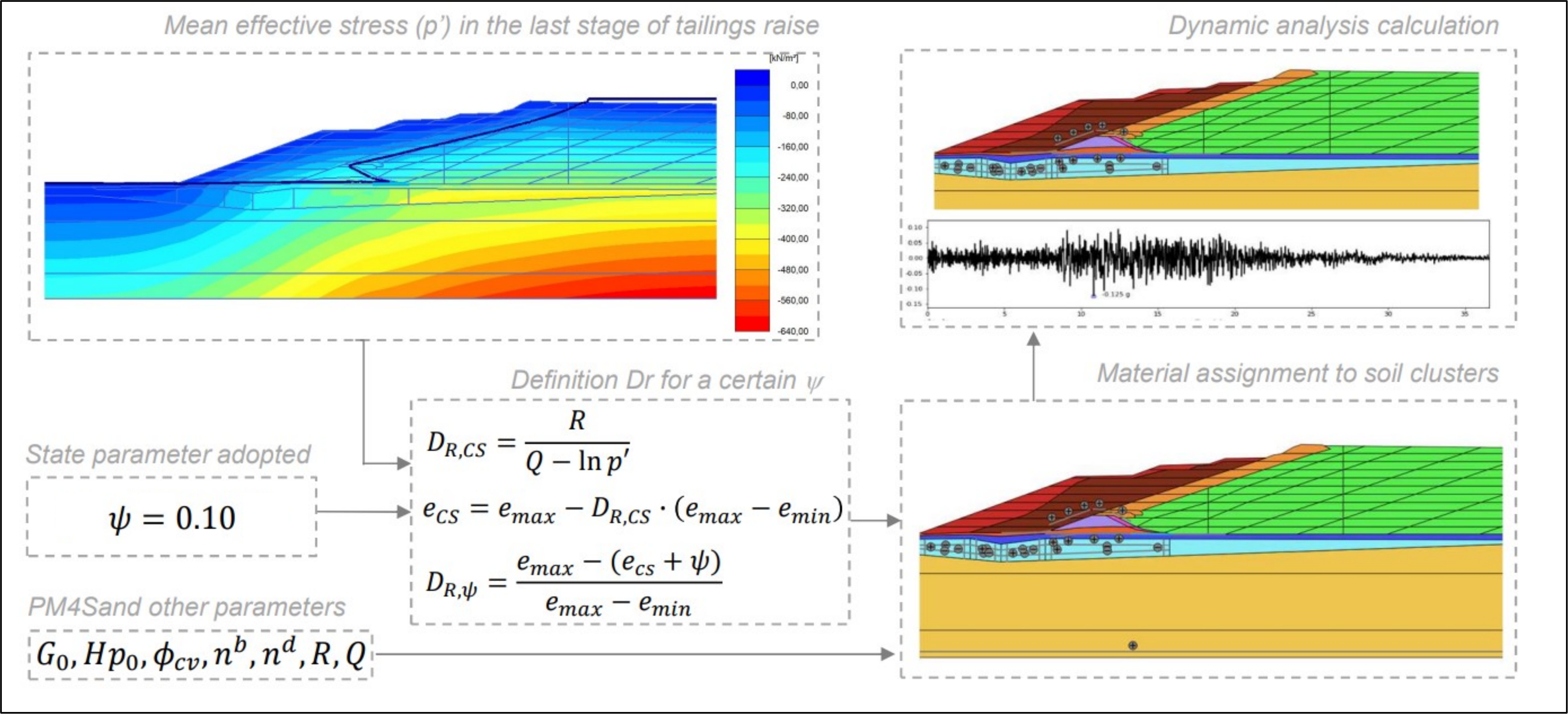}}
	\caption{State parameter adoption procedure for PM4Sand model.}
	\label{fig:PM4sand}
\end{figure}

The dynamic modelling starts from the last staged construction phase, representing the TSF crest at 130 m (i.e., the highest tailings level). For the PM4Sand model, the parameters are adopted following the procedure illustrated in Figure \ref{fig:PM4sand} and described as follows:
\begin{itemize}
	\item[•]	The average mean effective stress is obtained as an output from the last phase of the staged construction for different zones of the model.
	\item[•] The state parameter $\psi$ is adopted according to Table \ref{t:PM4sand}. In this work, $\psi = 0.1$.
	\item[•]	The relative density is estimated based on the average mean effective stress and the state parameter selected.
	\item[•]	The relative density and the other PM4Sand parameters defined in Table \ref{t:PM4sand} are assigned for each soil cluster.
\end{itemize}

\section{Results and correlations}
\label{section:Examples}

\begin{figure}
	{\includegraphics[height=8cm]{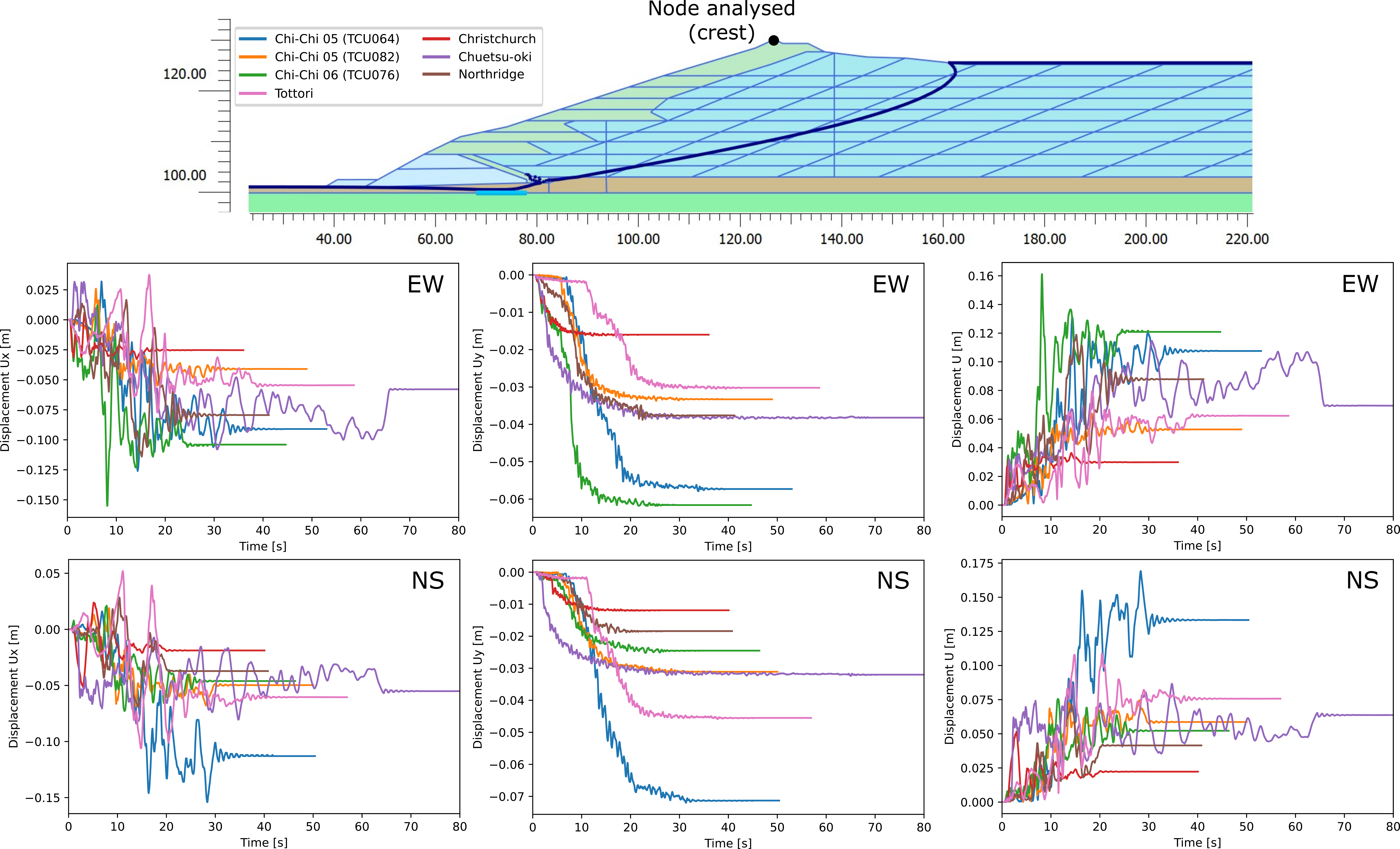}}
	\caption{Displacement evolution for a node at the crest of the TSF}
	\label{fig:Crest}
\end{figure}

\begin{figure}
	{\includegraphics[height=8cm]{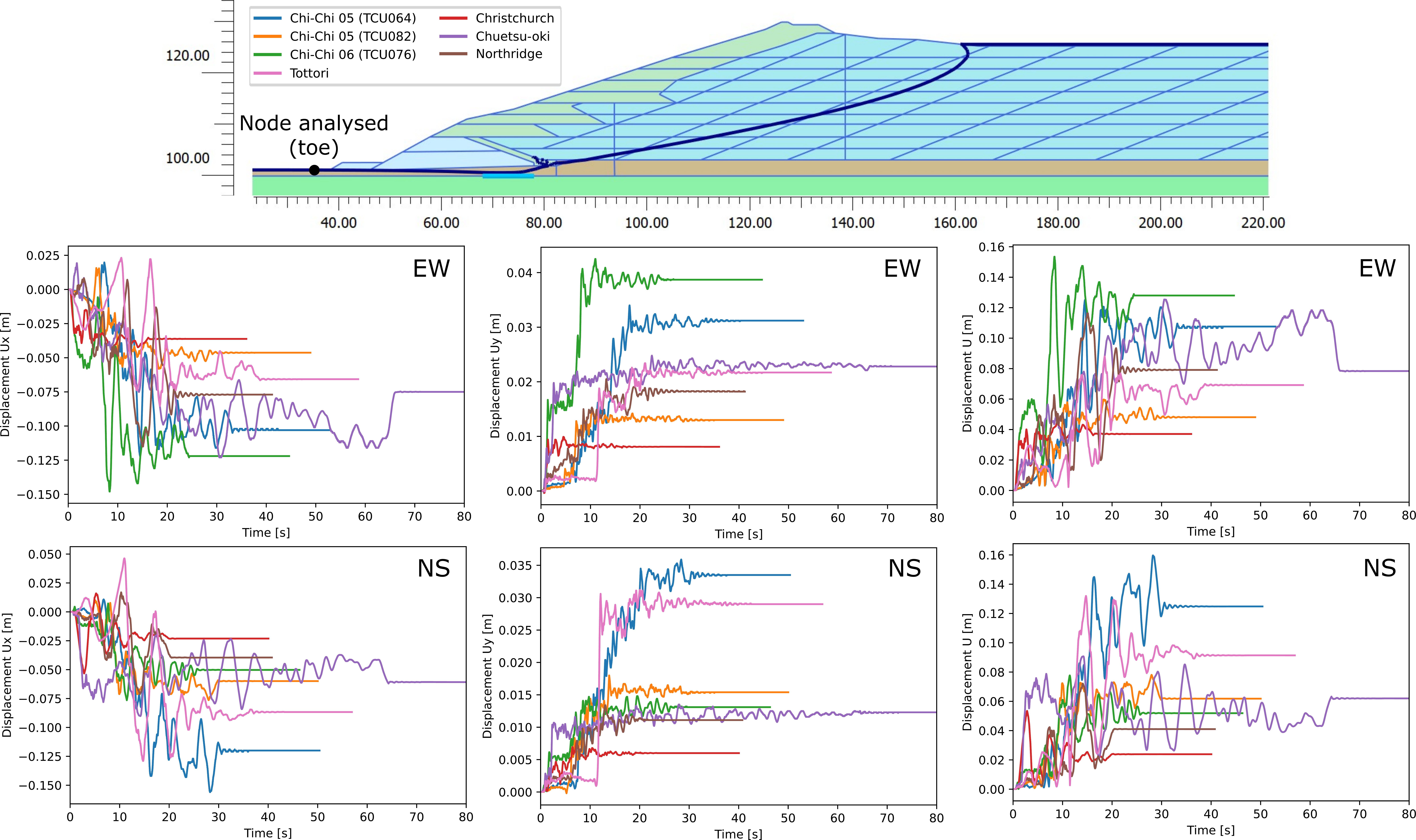}}
	\caption{Displacement evolution for a node at the toe of the TSF}
	\label{fig:toe}
\end{figure}

\begin{figure}
	{\includegraphics[height=21.5cm]{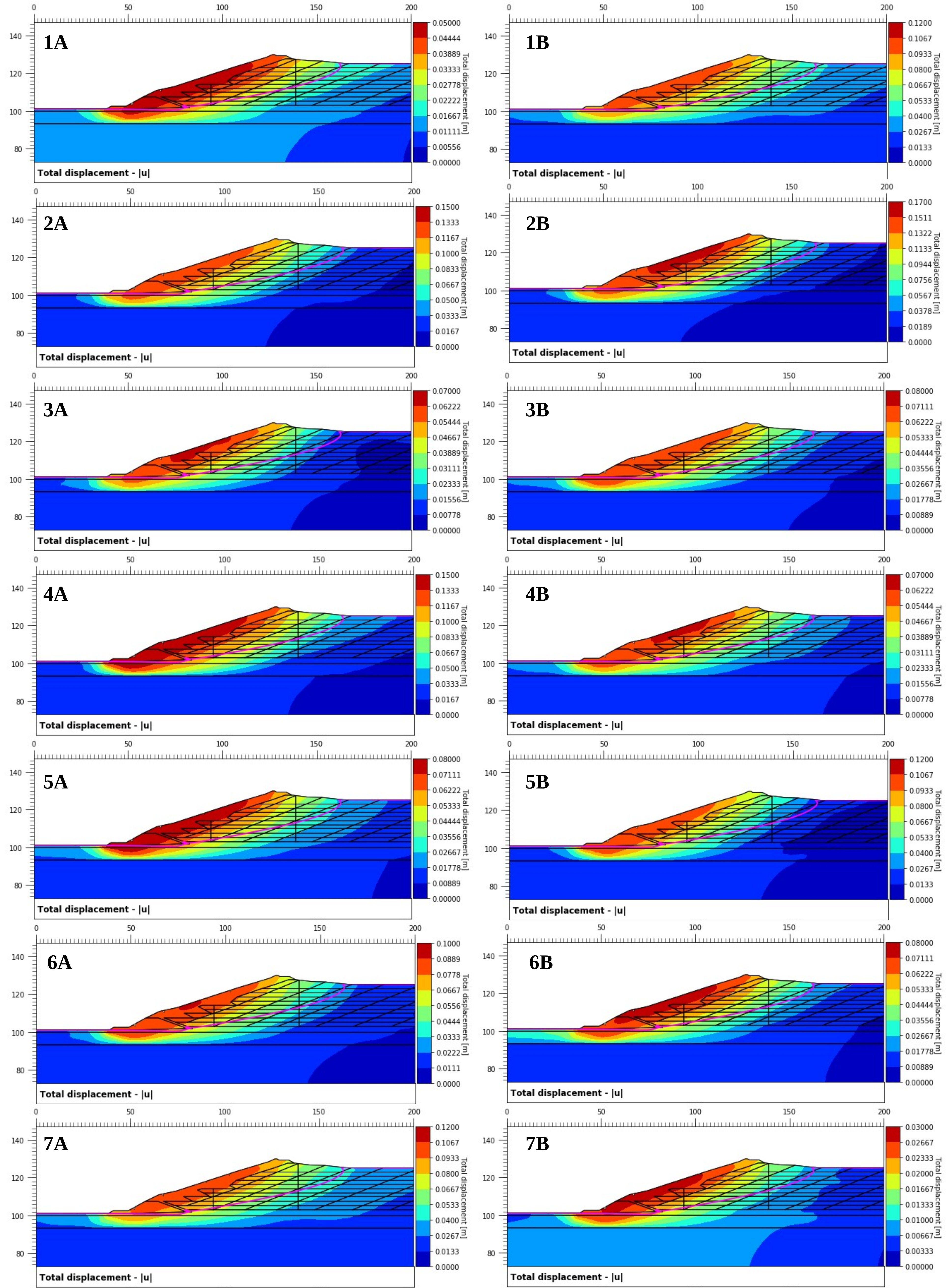}}
	\caption{Displacement contours at the end of the selected ground motions}
	\label{fig:dispcon}
\end{figure}

\begin{figure}
	{\includegraphics[height=21.5cm]{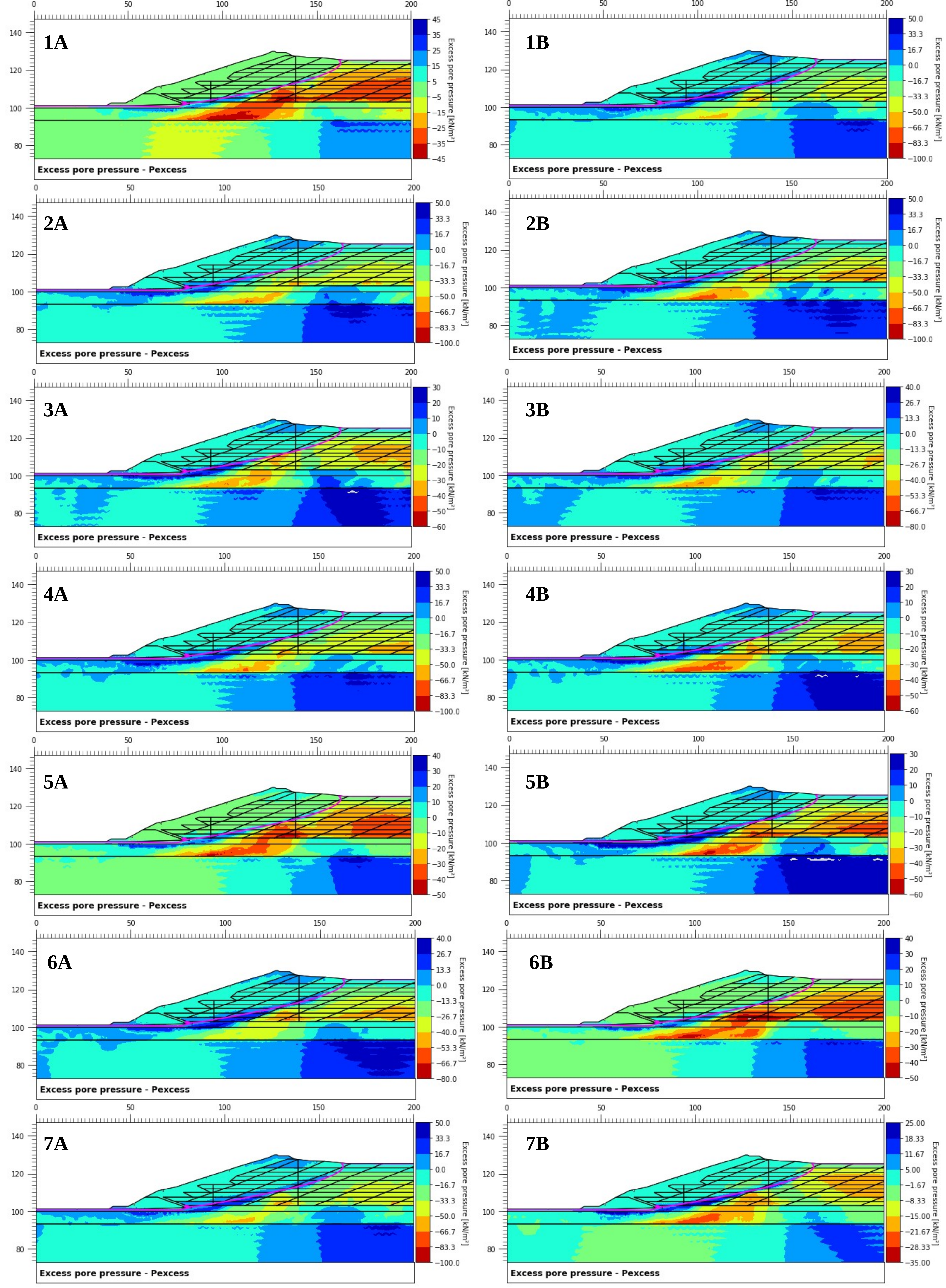}}
	\caption{Excess pore pressure contours at the end of the selected ground motions}
	\label{fig:porecon}
\end{figure}

Horizontal $Ux$, vertical $Uy$ and total displacements $U$ for the dynamic phases in two representative nodes of the dynamic modelling, crest and toe of the TSF, are displayed in Figure \ref{fig:Crest}, and Figure \ref{fig:toe}, respectively. Additionally, contours for vertical and total displacements and excess pore pressures at the end of each seismic simulation are depicted in Figures \ref{fig:dispcon} and \ref{fig:porecon}, respectively.
From the obtained results, the following observations can be deduced:
\begin{itemize}
	\item[•]	Maximum total displacements at the crest of the TSF reach values up to $160$ mm. In terms of the components, the vertical displacements (settlements) reach values up to $72$ mm, while the horizontal ones go up to $155$ mm. Maximum values are consistently associated with earthquakes 2A, 2B  and 4A, as was previously predicted by our IM.
	\item[•]	Excess pore pressures at the end of the seismic signal reach values up to 80 kPa. Pore pressure ratios fluctuate between 0.1 and 0.4, indicating a low risk of dynamic liquefaction. Typically, values around 0.9 are expected in high-risk liquefaction scenarios.
\end{itemize}

\begin{figure}
	\begin{center}
		{\includegraphics[height=3.25cm]{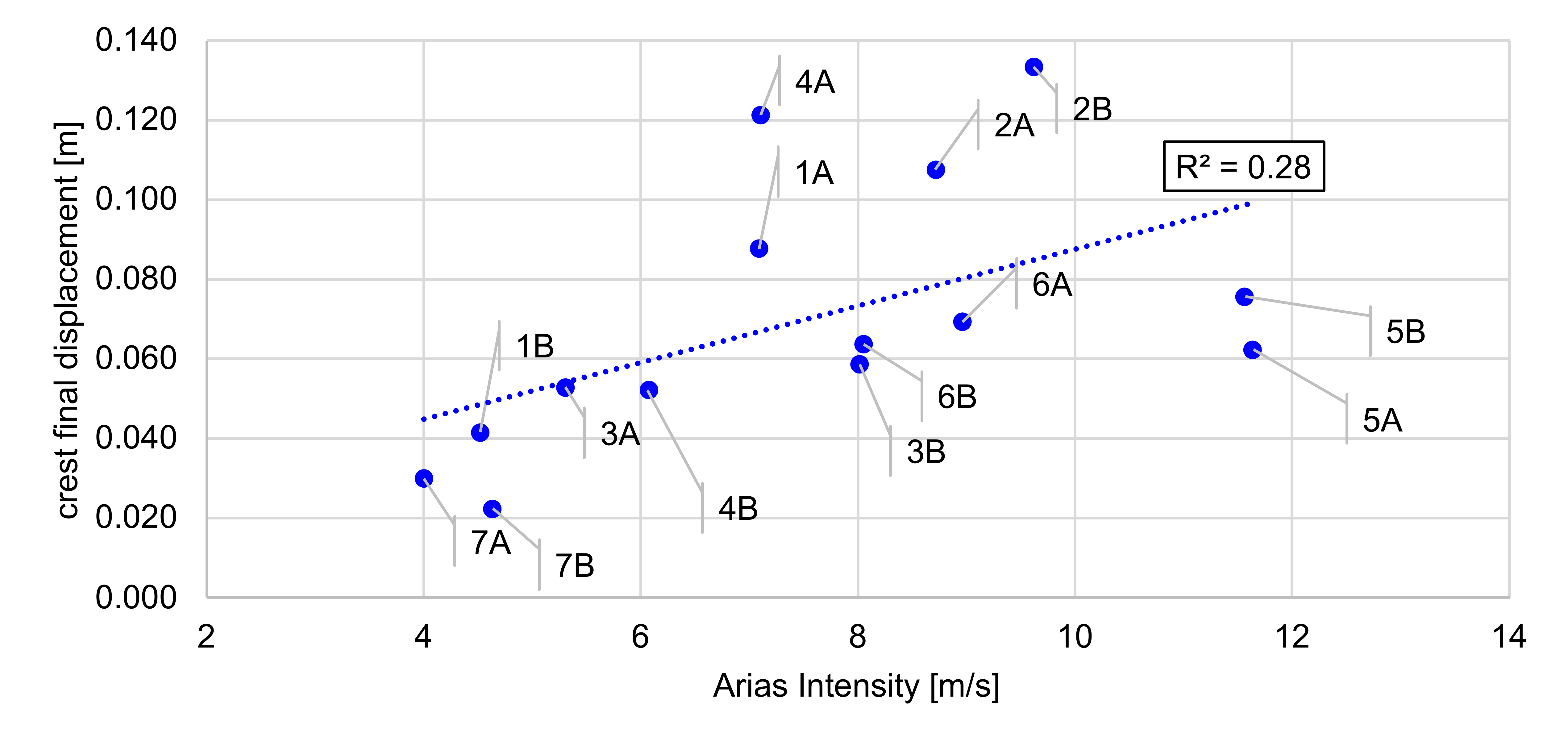}} 
		{\includegraphics[height=3.25cm]{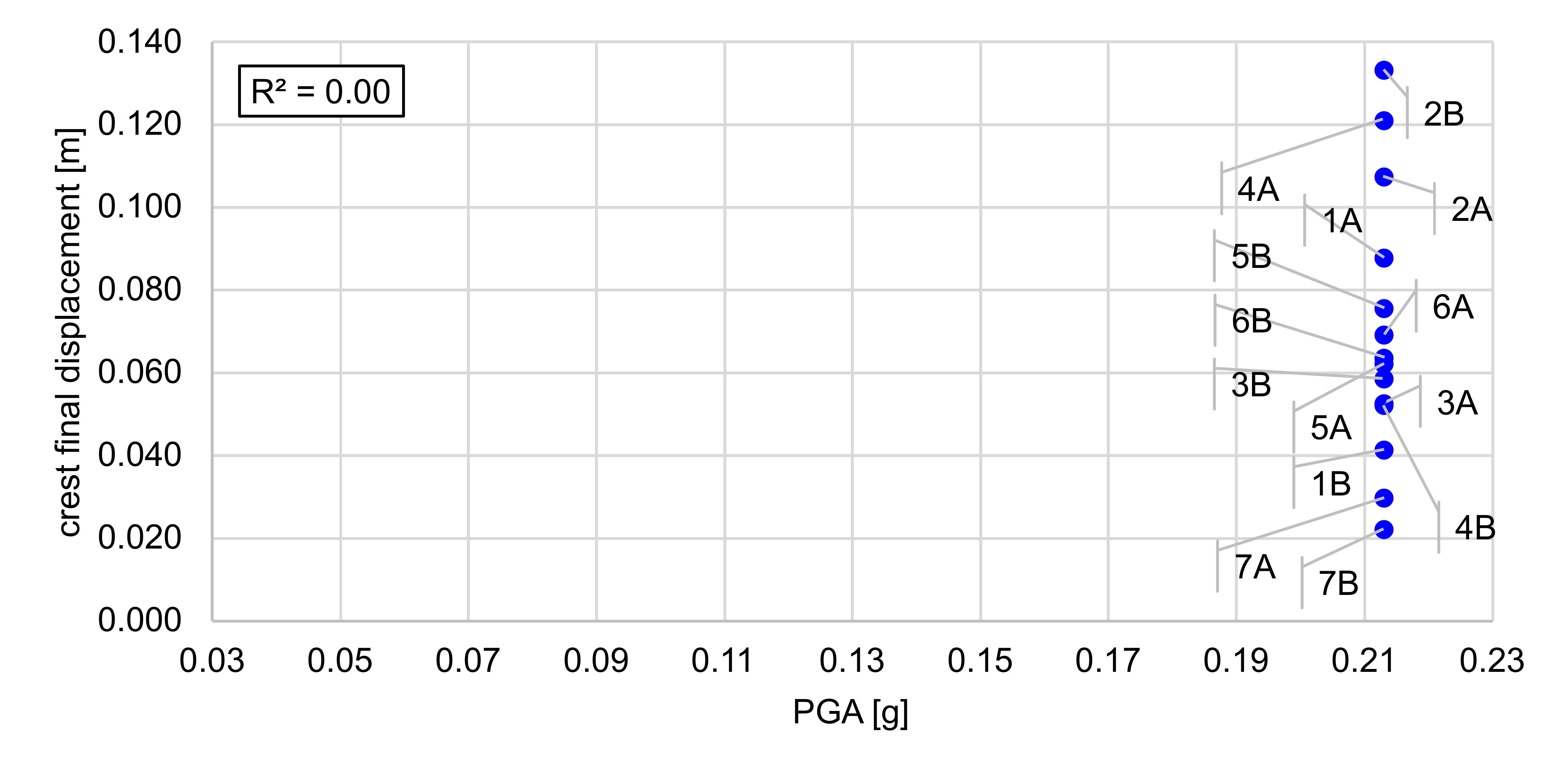}} \\
		{\includegraphics[height=3.25cm]{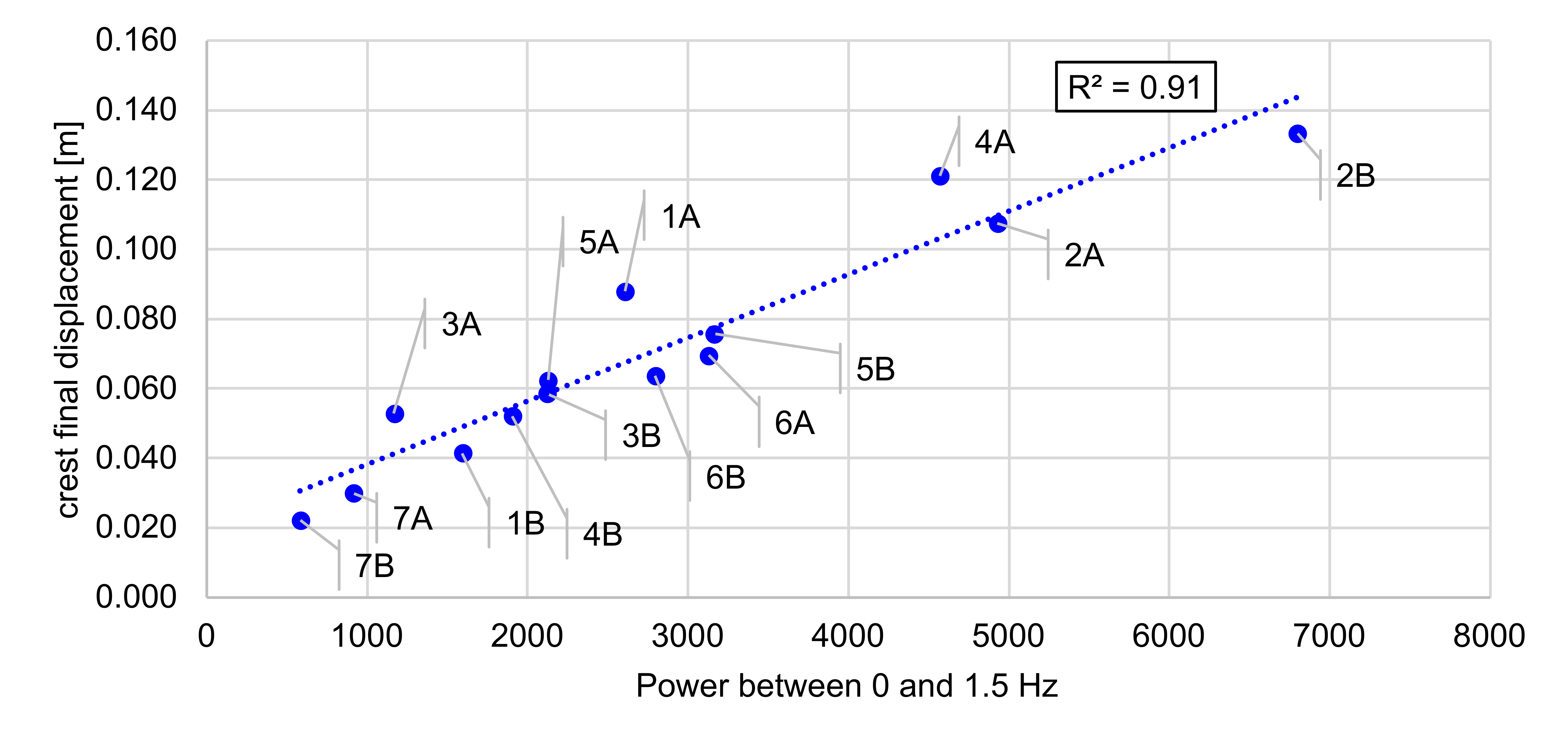}}
		{\includegraphics[height=3.25cm]{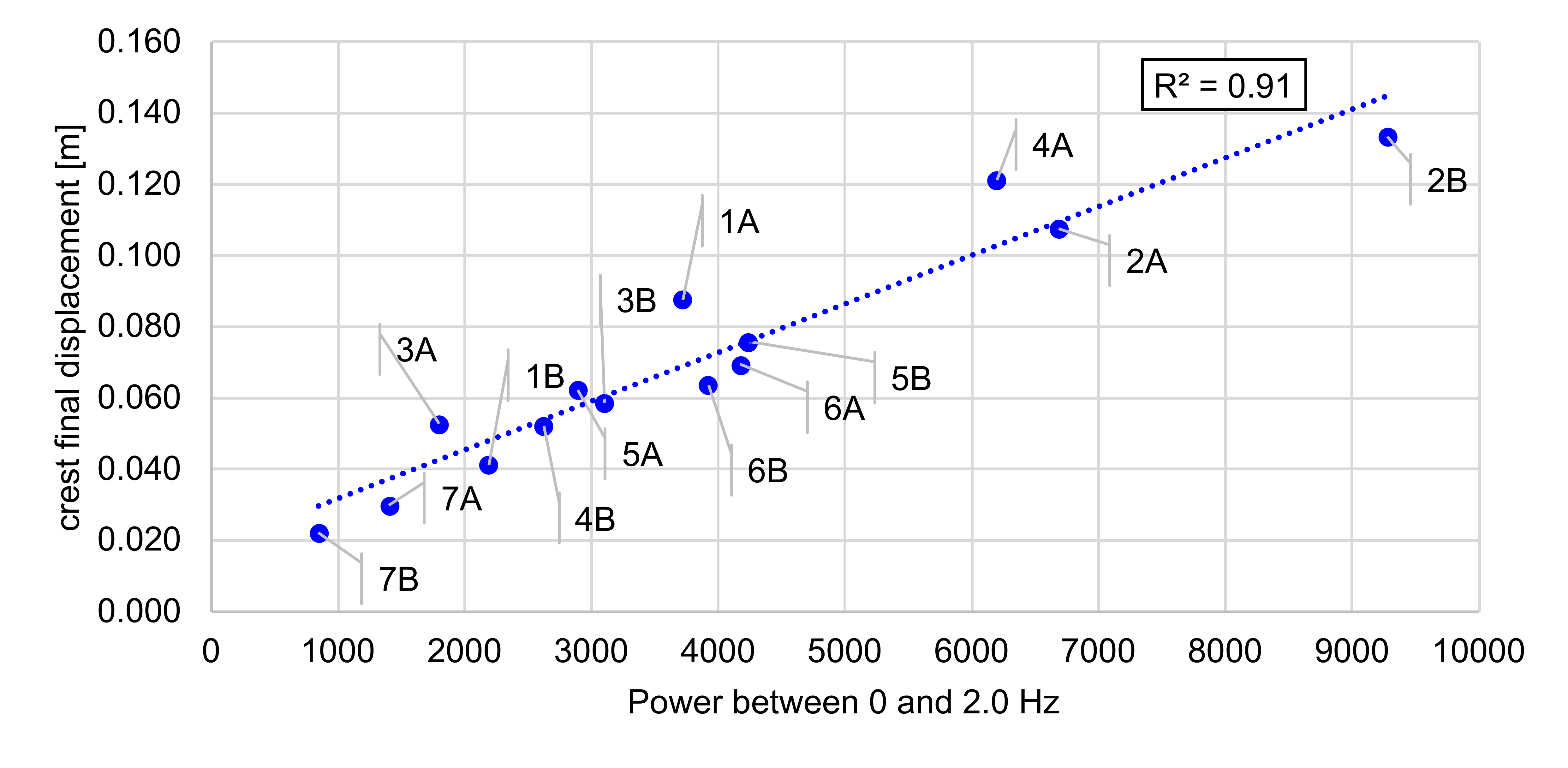}}\\
		{\includegraphics[height=3.25cm]{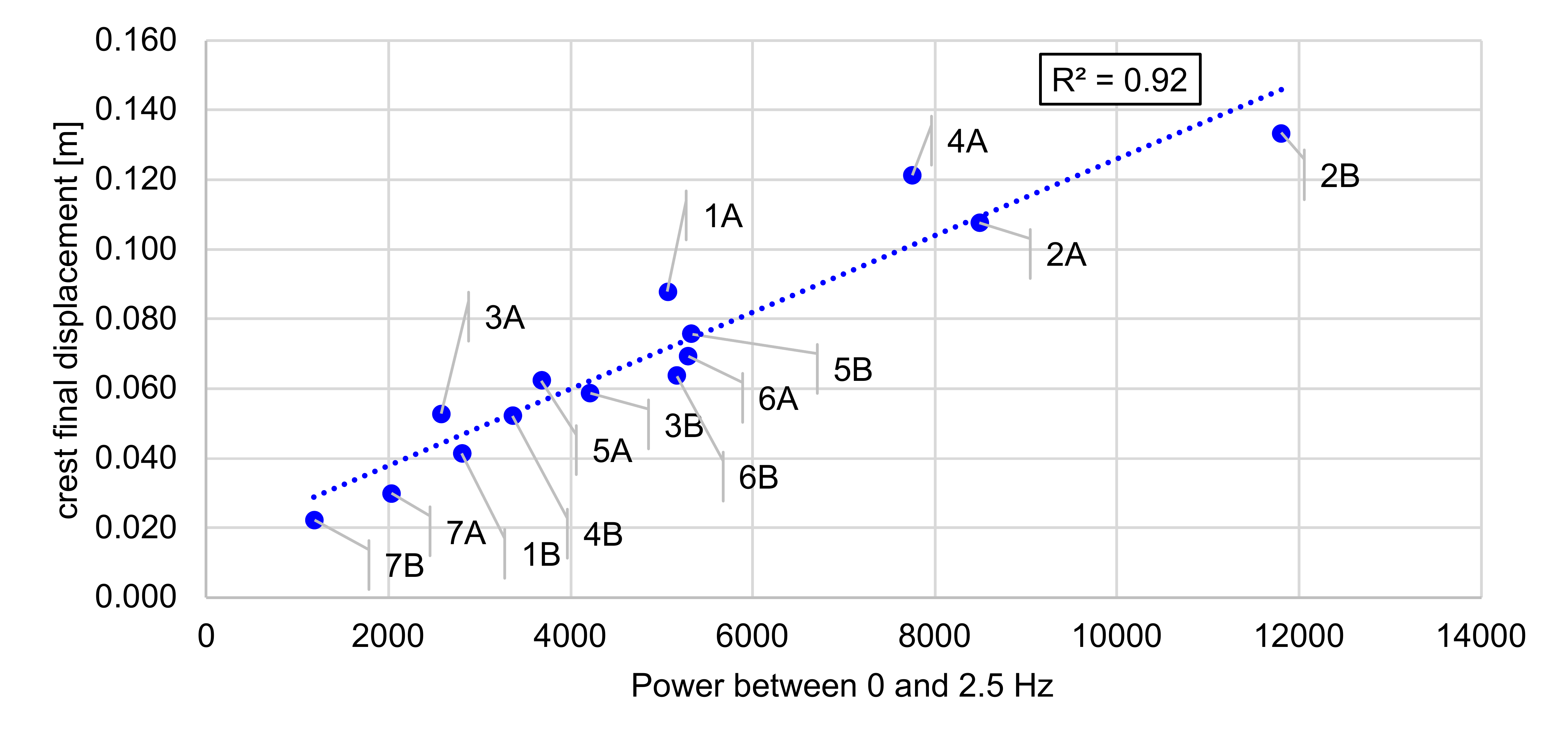}}
		{\includegraphics[height=3.25cm]{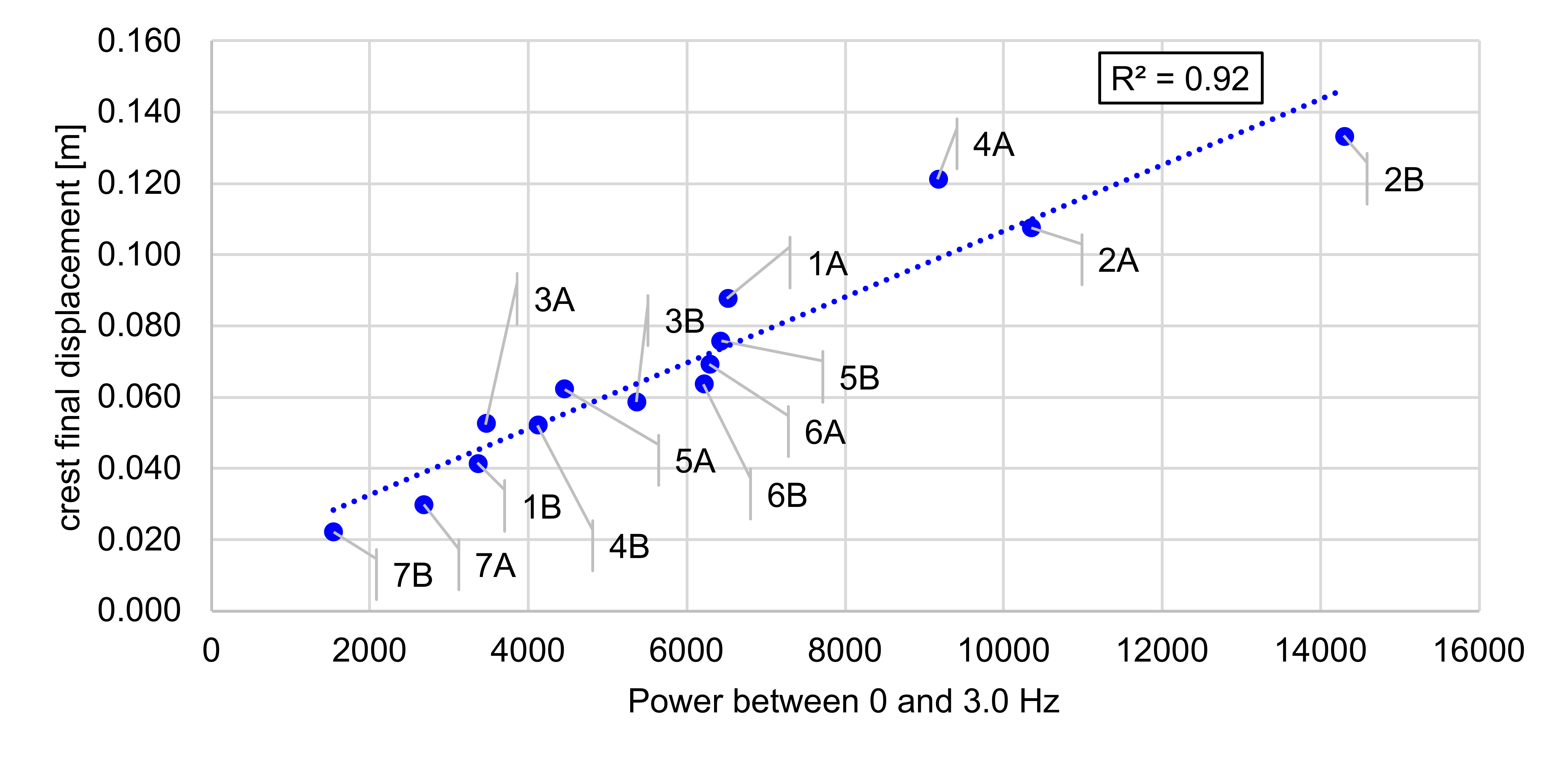}}
	\end{center}
	\caption{Residual displacement versus different intensity measures}
	\label{fig:FinalDisplacement}
\end{figure}

\begin{figure}
	\begin{center}
		{\includegraphics[height=3.25cm]{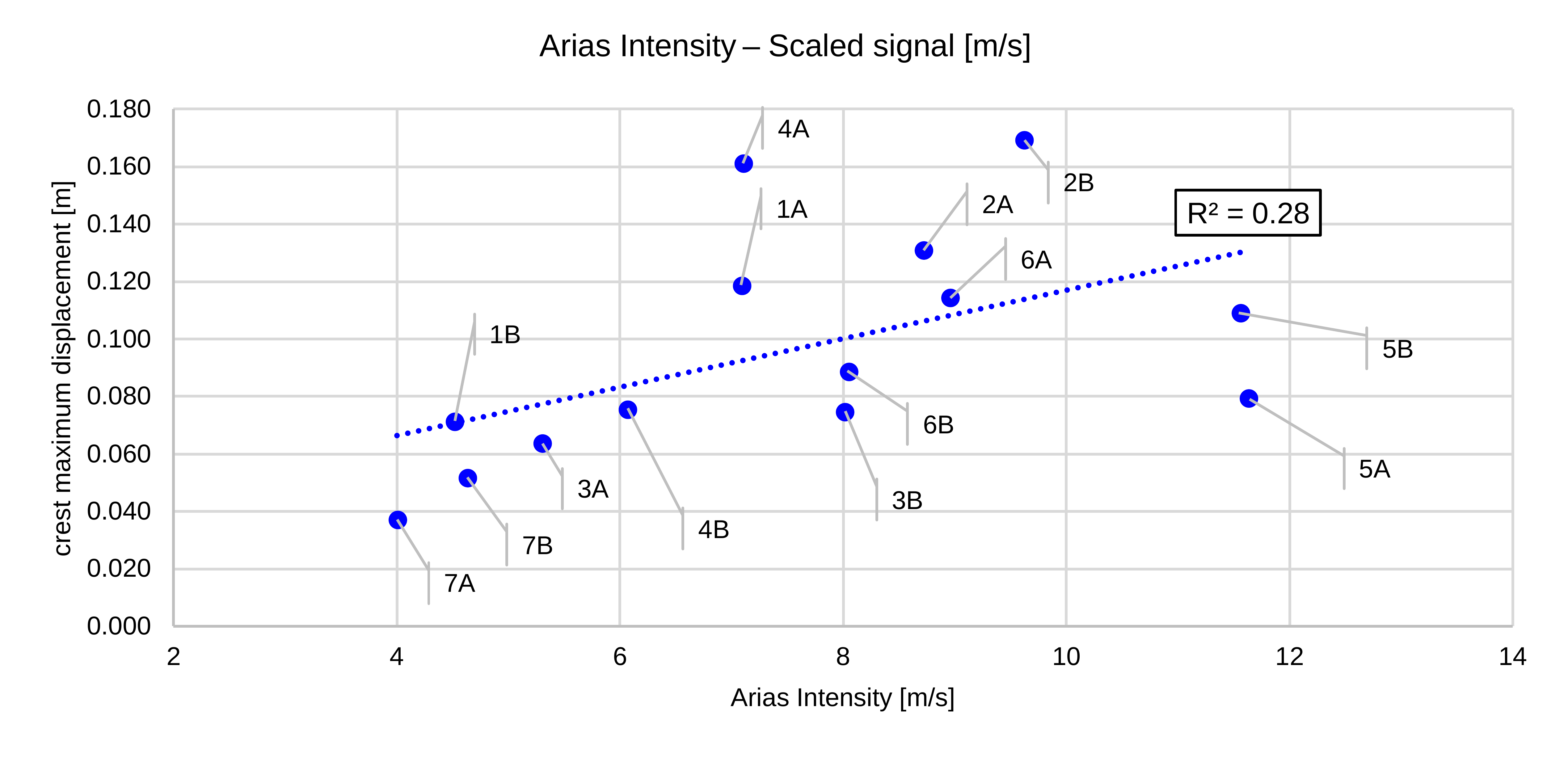}} 
		{\includegraphics[height=3.25cm]{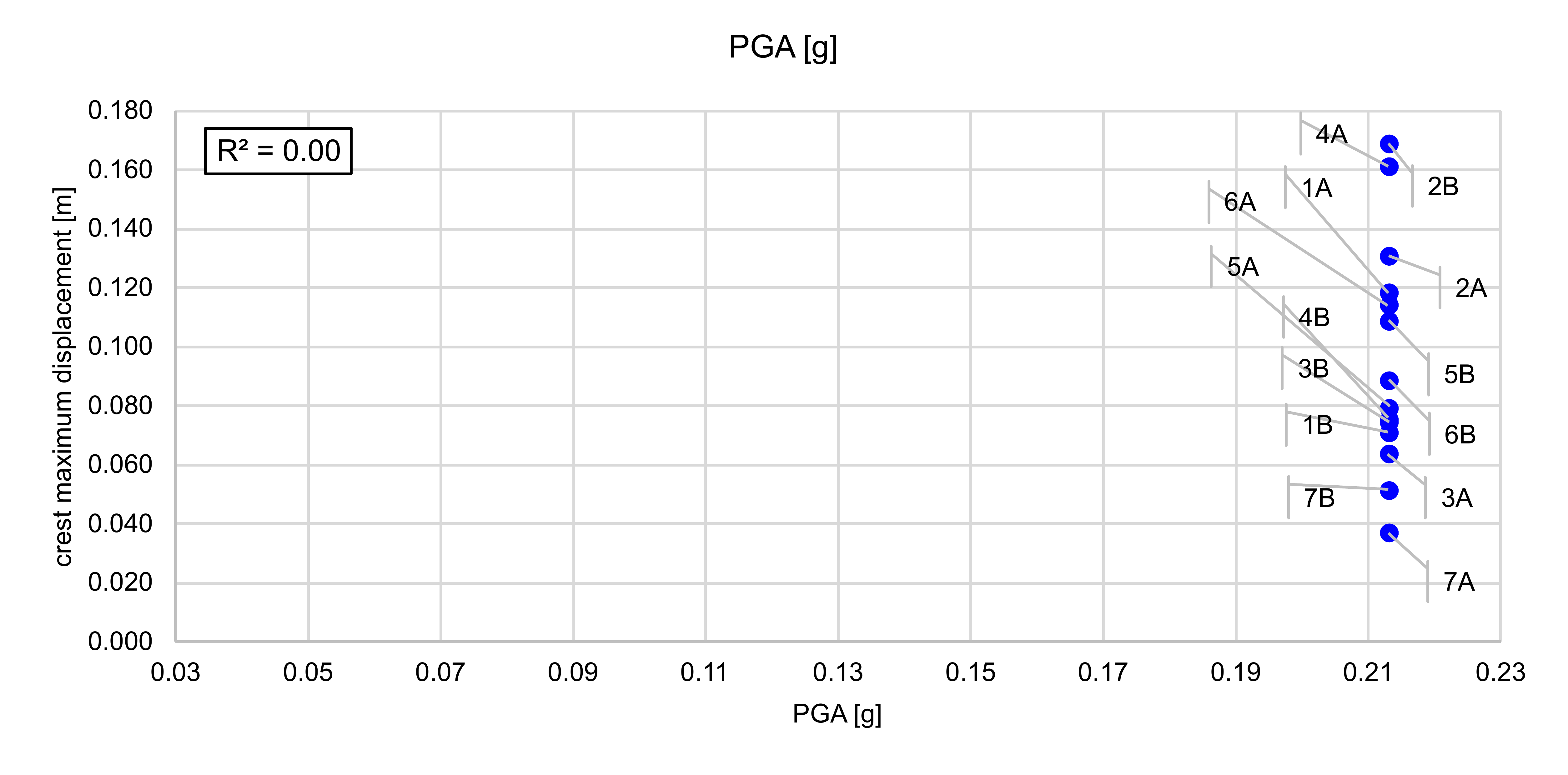}} \\
		{\includegraphics[height=3.25cm]{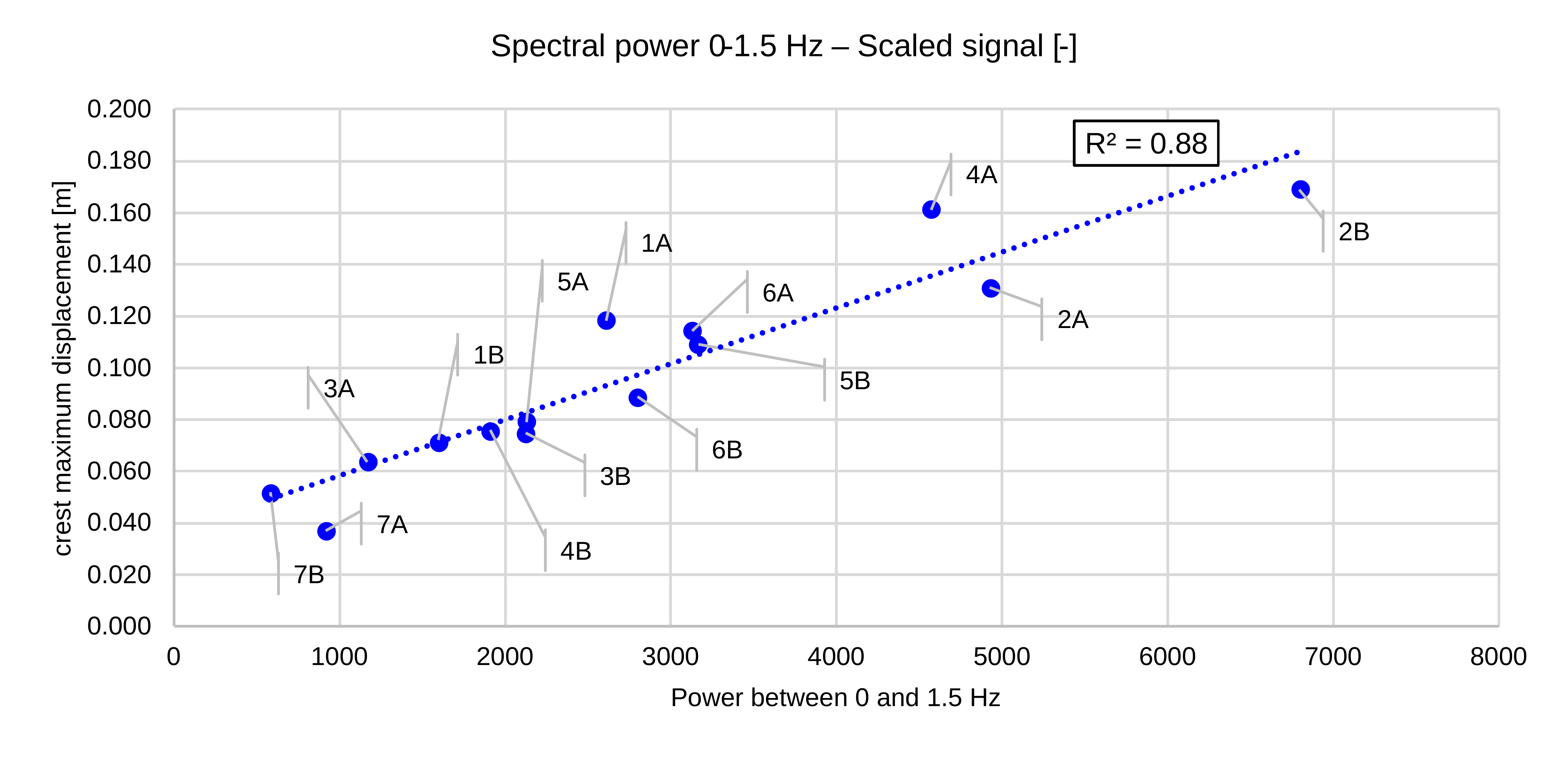}}
		{\includegraphics[height=3.25cm]{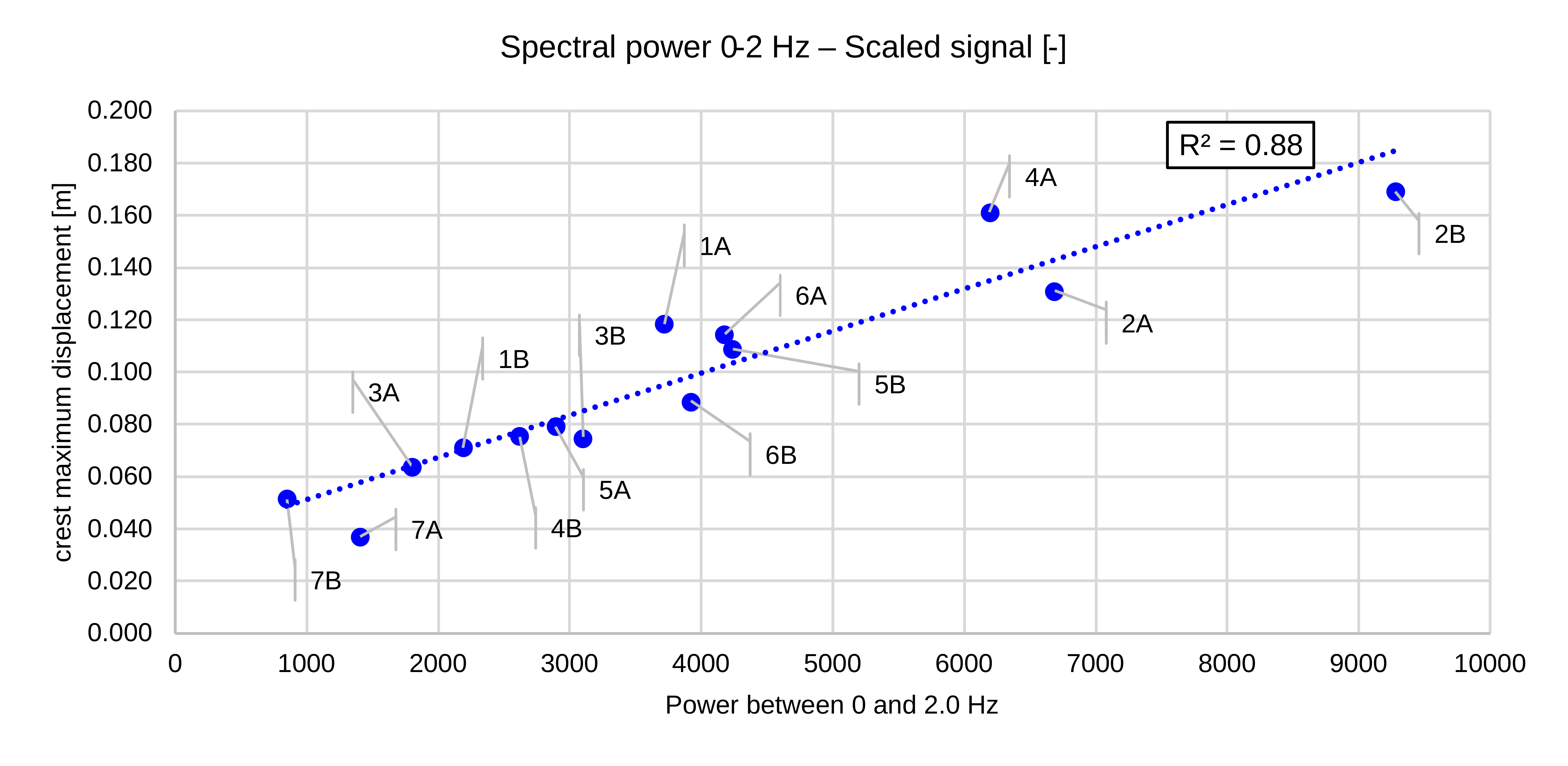}} \\
		{\includegraphics[height=3.25cm]{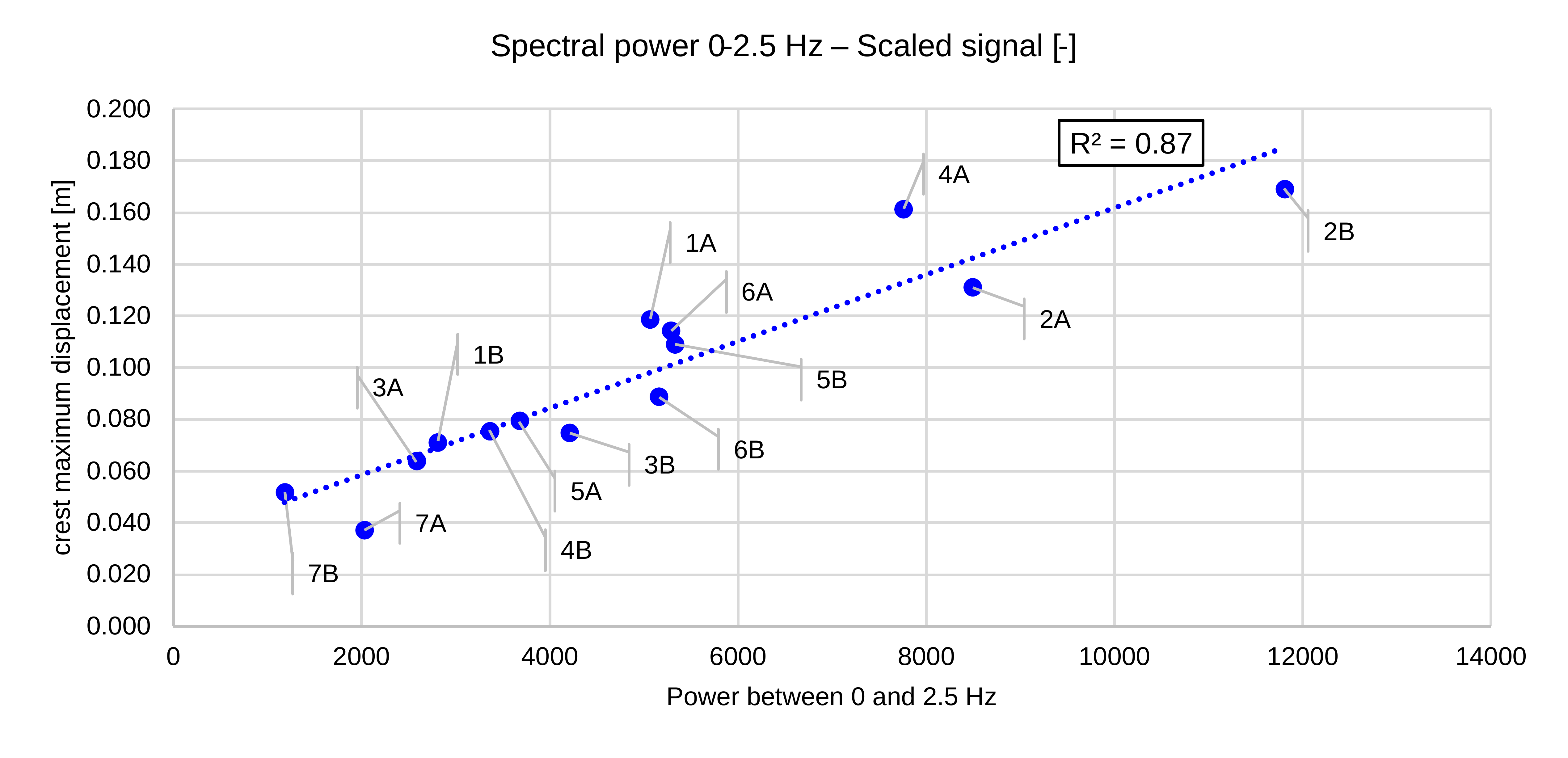}}
		{\includegraphics[height=3.25cm]{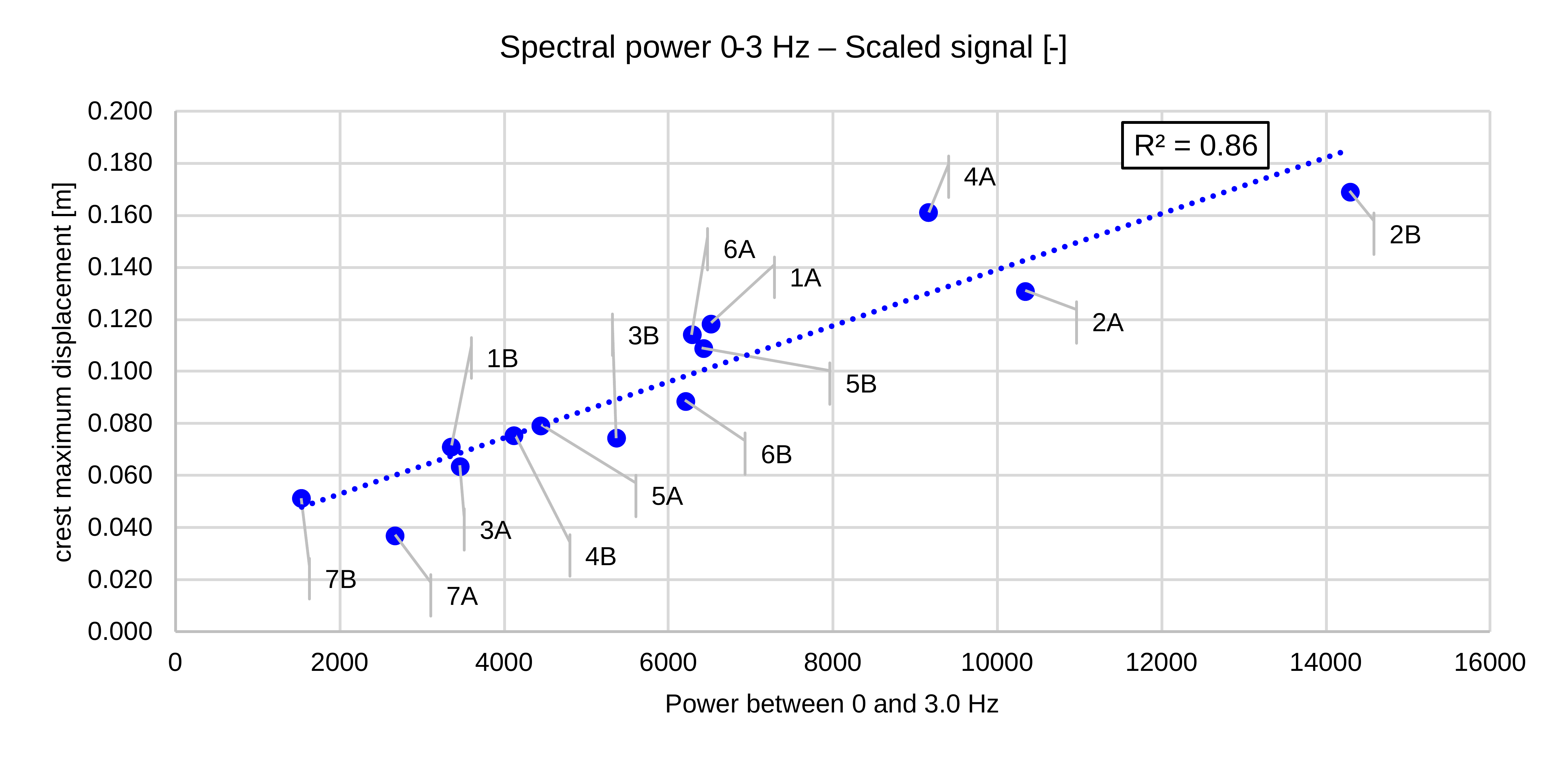}}
	\end{center}
	\caption{Maximum displacement versus different intensity measures}
	\label{fig:Finalmaxdispl}
\end{figure}

\begin{figure}
	\begin{center}
		{\includegraphics[height=3.25cm]{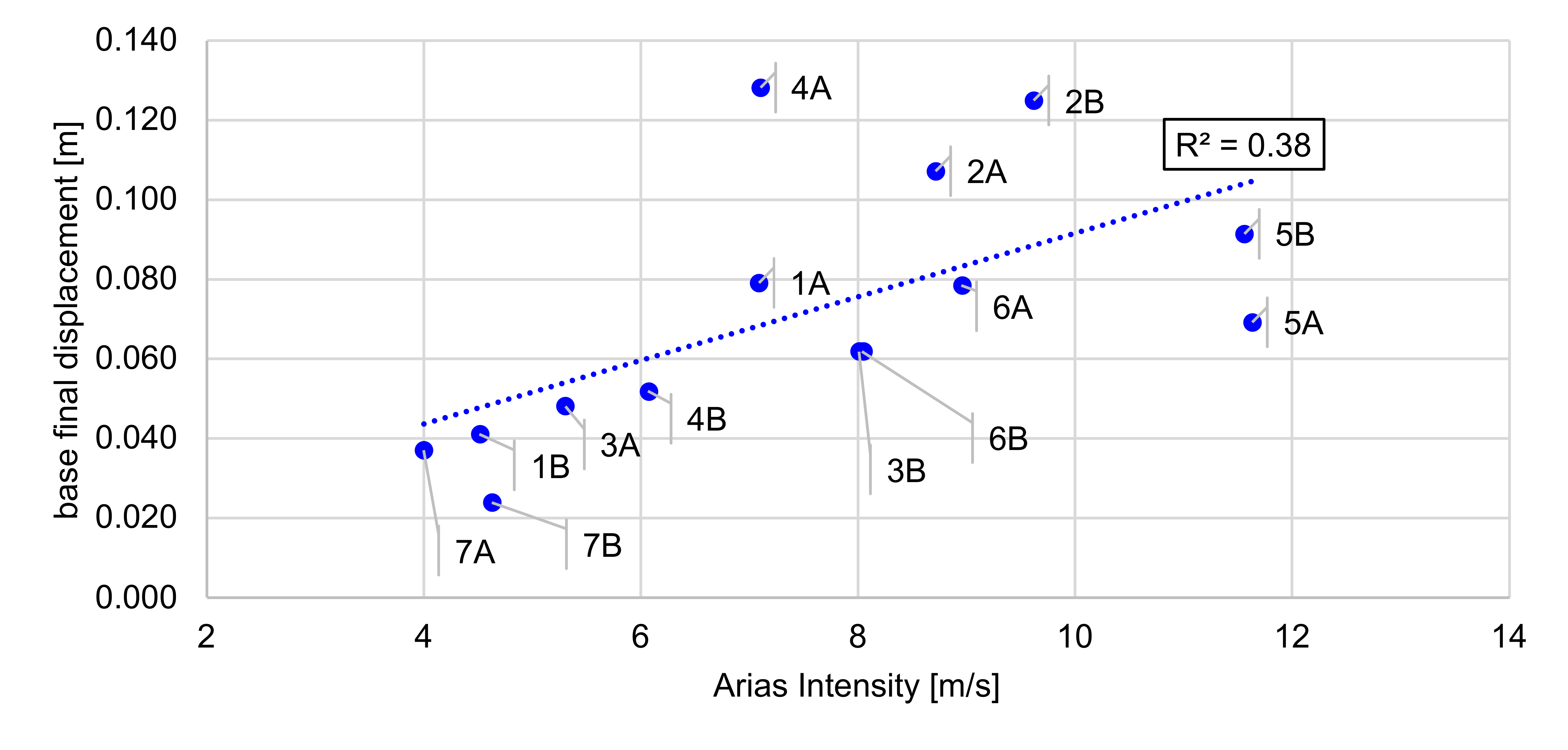}} 
		{\includegraphics[height=3.25cm]{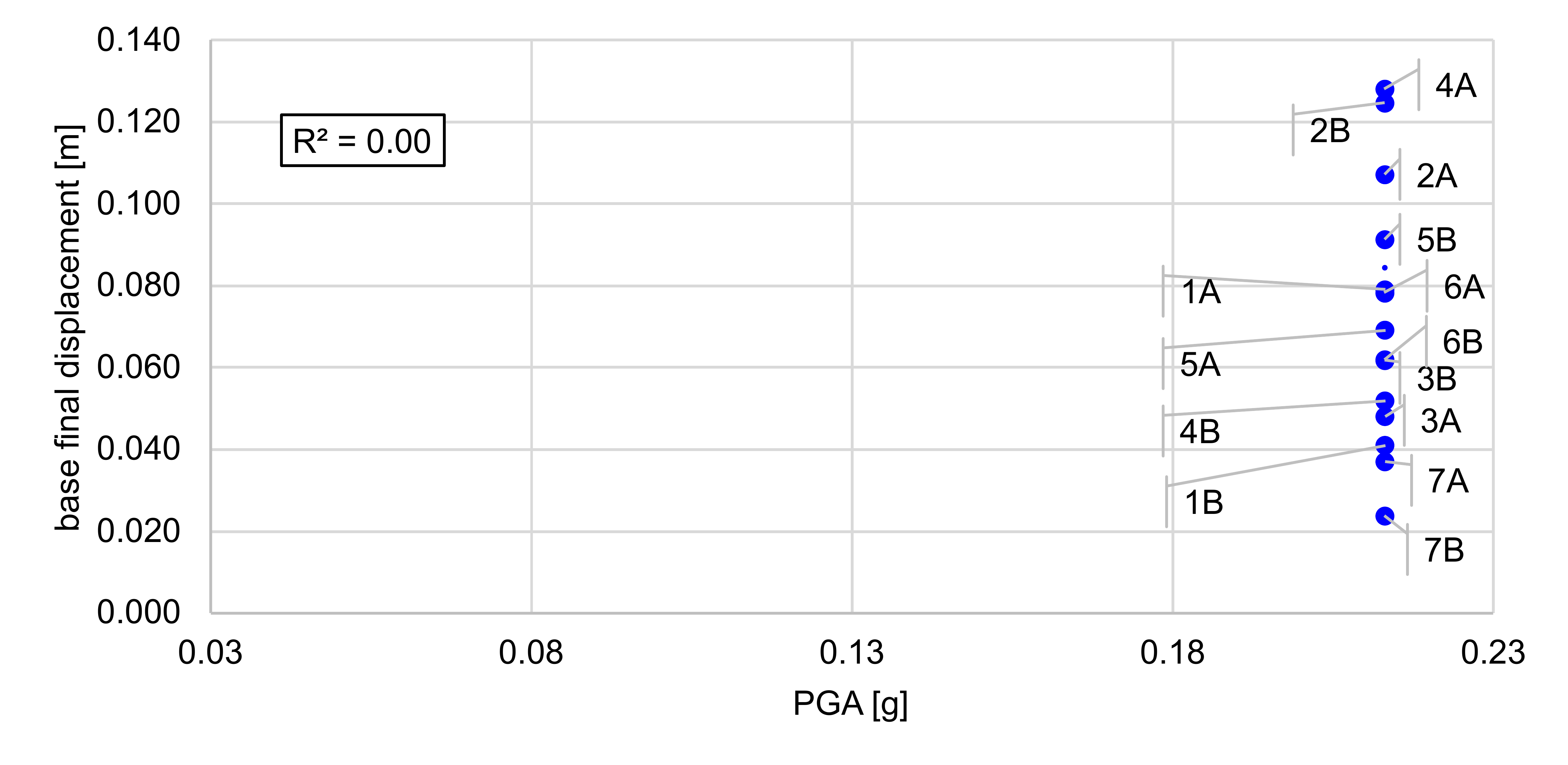}} \\
		{\includegraphics[height=3.25cm]{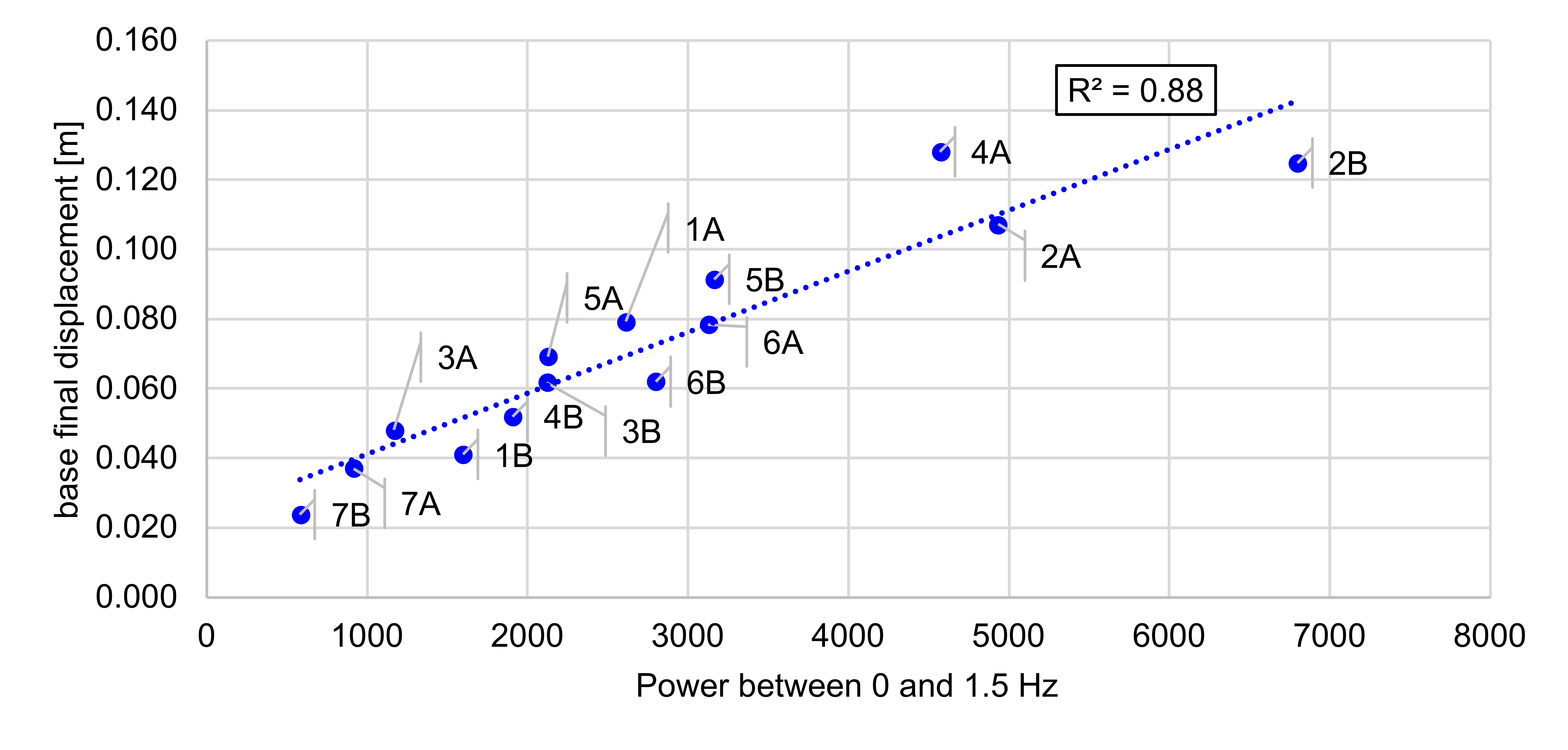}}
		{\includegraphics[height=3.25cm]{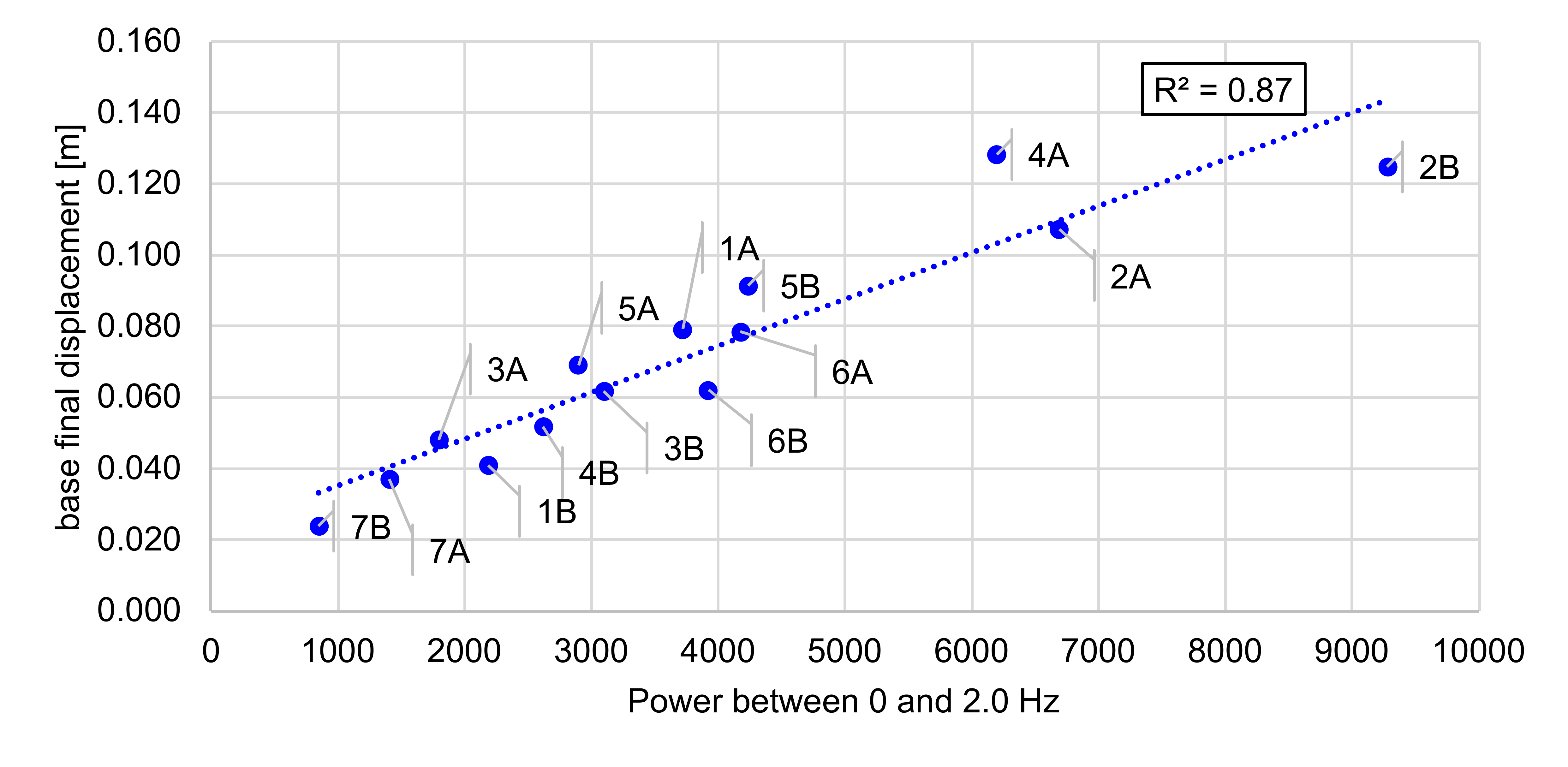}} \\
		{\includegraphics[height=3.25cm]{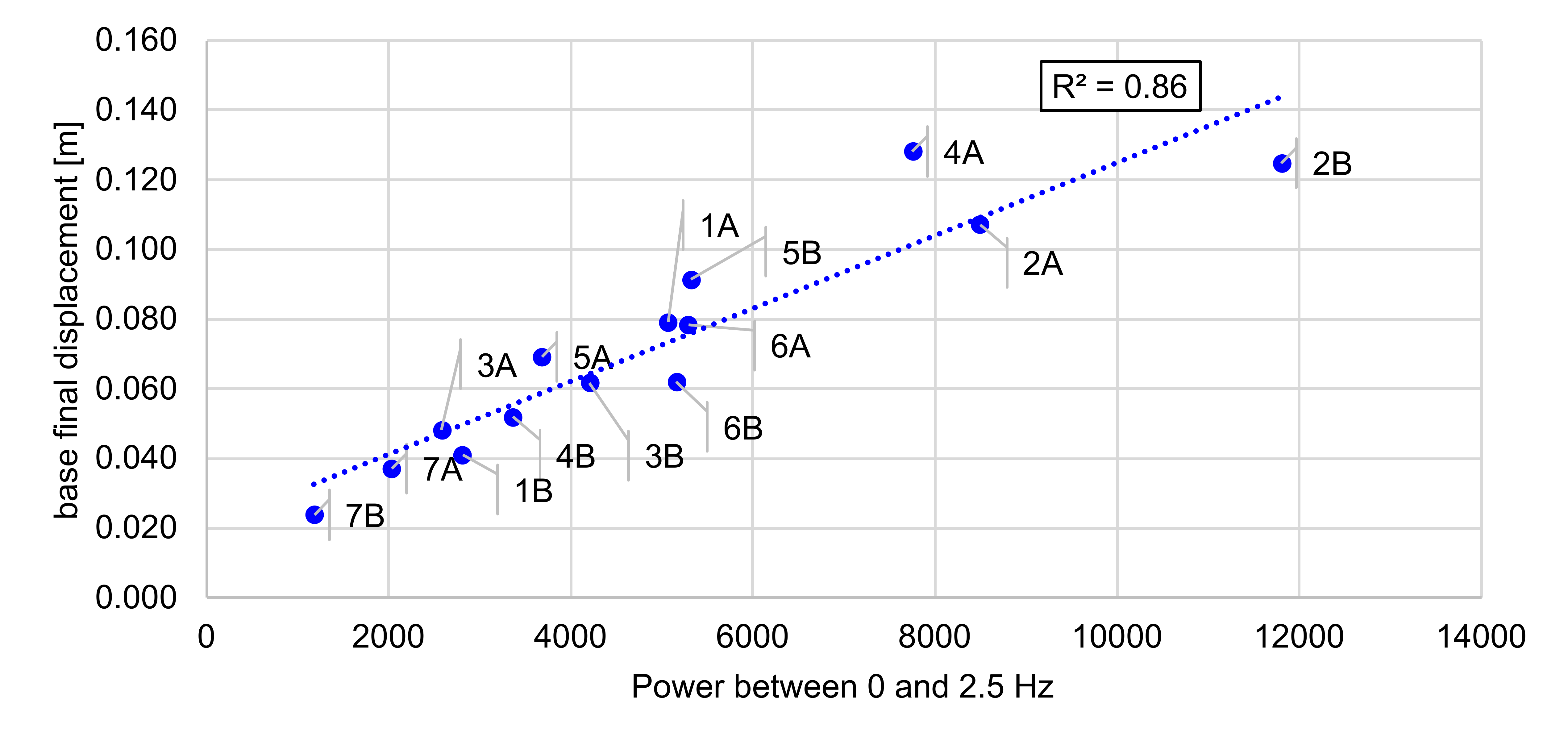}}
		{\includegraphics[height=3.25cm]{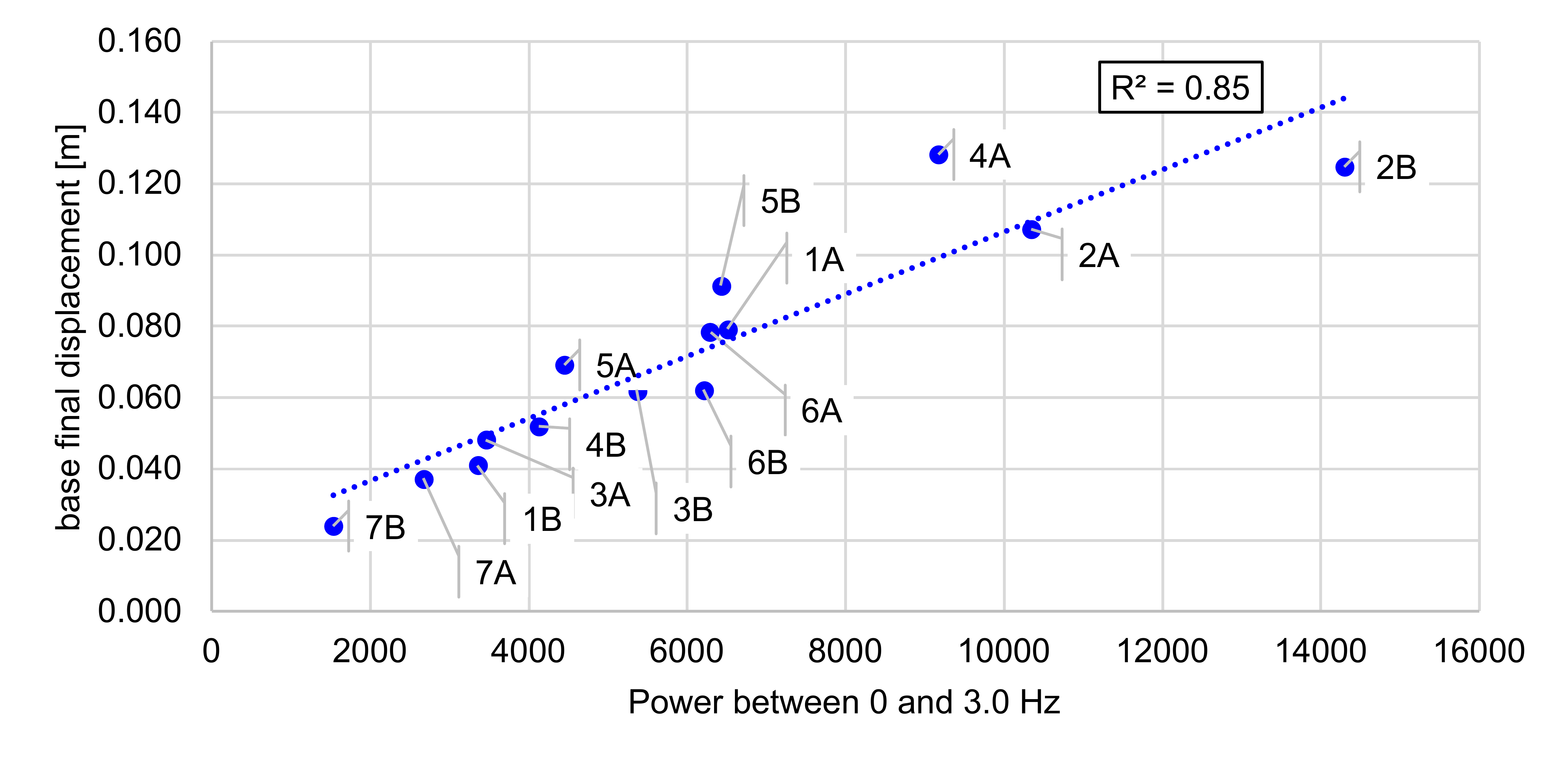}}
	\end{center}
	\caption{Final toe displacement versus different intensity measures}
	\label{fig:BaseDisplacement}
\end{figure}

Figures \ref{fig:FinalDisplacement},  \ref{fig:Finalmaxdispl} and \ref{fig:BaseDisplacement} show a series of different correlations between IMs and the residual displacements at the crest, the maximum displacement and the toe displacement of the TSF, respectively. From these, typical IMs such as PGA or Arias intensity correlate poorly with the final displacements, as the coefficients of determination are low (${R^2 < 0.3}$). However, IMs based on spectral power for a determined frequency window show a better correlation with the displacements, suggesting it is a better predictor of the seismic performance, as previously stated and proved in \cite{LABANDA2021106750} for high intense seismic signals. For these IMs, the coefficient of determination ${R^2}$ is above 0.85, representing a good fit of the regression model with the data. We can see that a widely used parameter in seismic hazard analysis, such as PGA, has a null capacity to predict the demand or residual displacement in this type of structure. These results also reinforced our previous statement that the seismic hazard analysis procedure must be reviewed and adapted for geotechnical structures, particularly tailings dams, because they are insensitive to higher frequencies energies and, therefore, the most demanding records are those with higher spectral power in lower frequencies.

\section{Conclusions}
\label{section:Conclusions}

In this paper, we proposed a new intensity measure (IM) based on spectral properties of the seismic record to assess tailings dams undergoing dynamic liquefaction. First, the mathematical definition of our IM was detailed and applied to a set of fourteen seismic records of mid-low intensity. Then, a model representing a section of the tailings storage facility (TSF) is described, including details regarding the mesh, constitutive model parameters, staged construction and its general layout. We subjected the model to these chosen ground motions and presented and discussed the series of time-history outcomes. We displayed correlation plots between maximum and residual crest displacement with residual base displacement and the proposed IMs. These showed that our IM has a prediction accuracy of around $R^2 = 0.90$, while a classical IM such as the Arias intensity is about $R^2 = 0.30$ with extreme cases such as the PGA case displaying null correlation. In conclusion, we validated our proposal and showed that it is a superior alternative for seismic signal selection when tailings dams are the structures under analysis. 


%



\bibliographystyle{unsrt}
\bibliography{mybibfile}

\end{document}